\documentclass[11pt]{article}              

\usepackage[margin=25mm]{geometry}
\usepackage{graphicx}
\usepackage{colortbl}
\usepackage{subfig}
\usepackage{float}
\usepackage{fancyhdr}
\usepackage{amssymb,amsfonts,amsmath,amsthm}
\usepackage{natbib}
\usepackage{bstnotations}
\usepackage[flushleft]{threeparttable}

\usepackage{authblk}
	
\graphicspath{{./},{./Figures/}}

\newcommand{\fxd}{f_{\ve{X}|\ve{d}}}
\newcommand{\fz}{f_{\ve{Z}}}
\newcommand{\fw}{f_{\ve{W}|\ve{d}}}
\newcommand{\Dd}{{\mathbb D}}
\newcommand{\cost}{{\mathfrak c}}
\begin{document}
\title{Surrogate-assisted reliability-based design optimization: a survey and a new general framework} 

\author[1]{M. Moustapha} \author[1]{B. Sudret}

\affil[1]{Chair of Risk, Safety and Uncertainty Quantification,
  
  ETH Zurich, Stefano-Franscini-Platz 5, 8093 Zurich, Switzerland}

\date{}
\maketitle

\abstract{Reliability-based design optimization (RBDO) is an active field of research with an ever increasing number of contributions. Numerous methods have been proposed for the solution of RBDO, a complex problem that combines optimization and reliability analysis. Classical approaches are based on approximation methods and have been classified in review papers. In this paper, we first review classical approaches based on approximation methods such as FORM, and also more recent methods that rely upon surrogate modelling and Monte Carlo simulation. We then propose a general framework for the solution of RBDO problems that includes three independent blocks, namely adaptive surrogate modelling, reliability analysis and optimization. These blocks are non-intrusive with respect to each other and can be plugged independently in the framework. After a discussion on numerical considerations that require attention for the framework to yield robust solutions to  various types of problems, the latter is applied to three examples (using two analytical functions and a finite element model).  Kriging and support vector machines together with their own active learning schemes are considered in the surrogate model block. In terms of  reliability analysis, the proposed framework is illustrated using both crude Monte Carlo and subset simulation. Finally, the covariance-matrix adaptation - evolution scheme (CMA-ES), a global search algorithm, or sequential quadratic programming (SQP), a local gradient-based method, are used in the optimization block. The comparison of the results to benchmark studies show the effectiveness and efficiency of the proposed framework. 
	% \\[1em] 

}

\maketitle

%%%%%%%%%%%%%%%%%%%%%%%%%%%%%%%%%%%%%%%%%%%%%%%%%%%%%%%%%%%%%%%%%%%%%%%%%%%%
\section{Introduction}
Realistic design of structures requires taking into account uncertainties which are ubiquitous to real world applications, \eg in manufacturing tolerances, loads or environmental and operational conditions. Casting the problem as a standard deterministic design does not allow one to safeguard the structure against unforeseen failures. Various methods have been introduced in the literature and applied by field engineers for the design of structures under uncertainties. \emph{Partial safety factors} \citep{Ditlevsen1996}, for instance, allow the analyst to implicitly account for uncertainties through the use of conservative characteristics  design parameters. However, such an approach often results in unnecessarily conservative and therefore costly design solutions. More convenient approaches have been developed under the framework of \emph{reliability-based design optimization}, where in most cases, the design constraints are assessed in terms of failure probabilities with respect to some predefined performance function. This implies modelling probabilistically the sources of uncertainty and then properly propagating those uncertainties to the quantities of interest. The complexity of this problem stems from the coupling of optimization and structural reliability analysis. A natural approach to its solution consists in nesting the two levels \ie exploring the design space using an appropriate optimization algorithm and computing the failure probability of each explored design using a structural reliability method. This is known in the literature as the \emph{two-level} approach \citep{Chateauneuf2008b} and includes popular methods such as the \emph{reliability index approach} \citep{Nikolaidis1988} or the \emph{performance measure approach} \citep{Tu1999}. In two-level approaches, the inner reliability analysis is carried out through an approximation method, typically the first-order reliability method (FORM). The cost of this naive approach can be somewhat high as it may require an overall large number of performance functions evaluations. Simplifying formulations such as the so-called \emph{decoupled} or \emph{mono-level} approaches have also been proposed in the literature (See \citet{Aoues2010, Valdebenito2010}). Despite some gain in efficiency, they remain expensive for the solution of real world problems, \ie they require thousands of computer model runs. Additionally, the use of approximation methods to evaluate the structural reliability constraints can hinder their application scope, due to their lack of convergence for highly non-linear performance functions. Some authors have proposed methods that rely on direct Monte Carlo simulation to evaluate the failure probability. The interest in such approaches has dramatically grown with the introduction of surrogate modelling in the RBDO framework. In a nutshell, the idea of surrogate modelling is to replace the original expensive-to-evaluate model by a cheap approximation that can be used for subsequent analyses. In this paper, a classification of such surrogate-assisted RBDO approaches is proposed based on the different ways the surrogate model is introduced in the RBDO framework. Upon reviewing various methods in the literature, a unified \emph{modular} and \emph{non-intrusive} framework is proposed. The modularity here means that the framework consists of three \emph{independent} blocks, namely design optimization, reliability analysis and surrogate modelling. By non-intrusiveness, it is meant here that the analyst can plug-in any method of his choice for each block without the need of changing anything else in the remaining blocks. Furthermore, the surrogate models are built adaptively so as to enhance the efficiency of the proposed framework. As an illustration, different configurations of each block are considered and applied to three problems of increasing complexity, which involve two analytical functions and a finite element model. More specifically, we consider:
\begin{itemize}
	\item support vector machines and Kriging as surrogate models;
	\item crude Monte Carlo and subset simulation as reliability methods and
	\item constrained covariance matrix adaptation - evolution scheme (CMA-ES) and sequential quadratic programming (SQP) as optimizers.
\end{itemize}

%%%%%%%%%%%%%%%%%%%%%%%%%%%%%%%%%%%%%%%%%%%%%%%%%%%%%%%%%%%%%%%%%%%%%%%%%%%%
\section{Formulation of the RBDO problem}
There are various ways of formulating a reliability-based design optimization problem. A common feature shared by all these approaches is that they attempt to trade the cost of the system with its reliability. \citet{Frangopol1985} and \citet{Chateauneuf2008b} make a short review from the historical viewpoint of the various RBDO formulations. The early approaches proposed to minimize a function which consists of an initial cost (mainly the addition of design and construction costs) and an expected cost of failure (See for instance \citet{Enevoldsen1994}). Due to the difficulty of properly assigning monetary cost to structural failure, this formulation was not of practical interest for engineering applications. It is now mostly investigated under the framework of \emph{risk-based} or \emph{life-cycle cost} optimization \citep{Beck2012, Frangopol2003}. Alternative formulations for RBDO have henceforth emerged. One, for instance, consists in maximizing the reliability, or equivalently minimizing the failure probability, under given cost constraints (See for instance \citet{Kuschel1997, Royset2001, Taflanidis2008}).

The approach we consider in this paper, which is also the prevalent in current research, consists in minimizing an initial cost under some probabilistic constraints \citep{Hilton1960}. Let $\ve{d} \in \Rr^{M_d}$ be an $M_d-$dimensional vector defining some design parameters of the structure. In the presence of uncertainties, the variability of these parameters can be described by introducing random variables $\ve{X} \in \Xx \subset \Rr^{M_d} \sim \fxd$. These variables are called \emph{design parameters with uncertainty}, or by slight abuse, \emph{random design variables} in the sequel. In some cases, design parameters are purely deterministic. To simplify the notation, they are equally considered as  $\ve{X}\prt{\ve{d}}$ at this stage, where the related random variables would simply have a zero variance. Other random parameters that may affect the structural response without being considered as decision parameters may also be of interest for the analysis. They are known as \emph{environmental} variables and are denoted here by $\ve{Z} \in \Zz \subset \Rr^{M_z} \sim \fz$. A typical RBDO formulation may then read \citep{Dubourg2011}:
\begin{equation}\label{eq:RBDOFormulation}
\begin{split}
& \ve{d}^\ast = \arg \min_{\ve{d} \in \Dd} \cost \prt{\ve{d}} \quad \text{subject to: }
\left\{ \begin{array}{ll}
\mathfrak{f}_j \prt{\ve{d}} \leq 0, \quad &  \acc{j = 1 \enum n_s}, \\
\Prob{\mathfrak{g}_k \prt{\ve{X}\prt{\ve{d}},\ve{Z}} \leq 0} \leq \bar{P}_{f_k}, \quad & \acc{k = 1 \enum n_h},
\end{array} \right.
\end{split}
\end{equation}
where $\cost$ is the cost function to be minimized under constraints that are classified into two categories. The first, $\mathfrak{f}_j, j = \acc{1 \enum n_s}$, are a set of deterministic analytical functions that define the \emph{feasible domain}, \eg bounds between the design parameters. The second category, which consists of probabilistic constraints, set an upper threshold on the failure probability for each identified failure mode of the structure (herein $\bar{P}_{f_k}, k=\acc{1 \enum n_h}$), assuming a series system behaviour. The latter is defined through a so-called \emph{limit-state function} which partitions the random input space into safe and failure domains. By introducing a vector $\ve{W} = \acc{\ve{X}\prt{\ve{d}},\ve{Z}}^T$ that gathers the random design and environmental variables, the failure probability for a given design $\ve{d}$ can be expressed as \citep{Ditlevsen1996}:
\begin{equation}\label{eq:Pf}
P_{f_k}\prt{\ve{d}} = \Prob{\mathfrak{g}_k\prt{\ve{W}\prt{\ve{d}} \leq 0}} = \int_{\mathcal{D}_f} \fw\prt{\ve{w}} d\ve{w},
\end{equation} 
where $\mathcal{D}_f = \acc{\ve{w} \in \Xx \times \Zz: \mathfrak{g}_k\prt{\ve{w}} \leq 0}$ represents the failure domain and $\fw$ is the joint distribution of all variables given a particular value of $\ve{d}$. 

Equation~\eqref{eq:Pf} constitutes a multidimensional integration problem over an implicitly defined domain whose solution is difficult. This difficulty to some extent explains the delay between the early RBDO formulations and the implementation of solution procedures which flourished in the 80s. In general, the reliability techniques used to solve this integration problem mainly rely either on first-order approximations or on simulation methods, as detailed in the sequel.   
%%%%%%%%%%%%%%%%%%%%%%%%%%%%%%%%%%%%%%%%%%%%%%%%%%%%%%%%%%%%%%%%%%%%%%%%%%%%

\section{A short review of RBDO solution techniques}

\subsection{Classical RBDO methods}
Reliability-based design optimization is a rather rich and active field of research experiencing a continuous flow of publications. All the proposed methods seek to reduce the computational cost of the RBDO problem, mostly by introducing approximations in the reliability analysis or by reformulating the optimization problem. We start our survey  using the classification suggested in \citet{Chateauneuf2008b, Aoues2010}, namely \emph{two-level}, \emph{mono-level} and \emph{decoupled} approaches. Some techniques do not entirely fit this classification though, as will be shown in the sequel. We then proceed with other, more recent approaches that used advanced techniques. Older reviews of RBDO can also be found in \citet{Valdebenito2010, Aoues2010}.

\subsubsection{Two-level approaches}
A two-level approach is the most direct way to solve a RBDO problem. It consists of two nested loops: the outer loop explores the design space via a suitable optimization scheme while the inner loop performs a reliability analysis. As the values of design parameters vary across iterations, it is necessary to repeatedly run a full reliability analysis, which may be computationally expensive. The overall cost of this approach is in most cases prohibitive. To alleviate this burden, the two main strategies in this category resort to approximation techniques, more specifically the first-order reliability method (FORM). The \emph{reliability index approach} (RIA) consists in simply using FORM in the inner loop while replacing the probabilistic constraint by an equivalent reliability index \citep{Nikolaidis1988, Lee2002}. An alternative formulation, the so-called \emph{performance measure approach} (PMA), relies on an inverse FORM analysis in the inner loop that transforms the probabilistic constraints into equivalent target performance measures \citep{Tu1999,Youn2005}. The main advantages of these approaches lie on the attractivity of FORM which, besides being relatively cheap, allows for an efficient computation of the gradient of the reliability index with respect to the design variables. However, FORM is also the weakness of these approaches as it shows limitations when it comes to problems involving either highly non-linear limit-state functions or multiple failure regions.     

Using the same framework for two-level approaches, some contributions have suggested the use of simulation methods in the inner loop (See for instance \citet{Royset2004}). Even when using advanced simulation methods whose cost is relatively small compared to crude Monte Carlo sampling, the overall cost of such approaches is prohibitively high. In general, they are coupled with surrogate models. This aspect is discussed in details later.

\subsubsection{Mono-level approaches}
Mono-level approaches attempt to solve the RBDO problem by avoiding the reliability analysis and enforcing optimality conditions as initially proposed in \citet{Madsen1992}. This, in theory, allows one to reduce the computational cost since the failure probability is no more explicitly computed. \citet{Kuschel1997} introduced the Karush-Kuhn-Tucker (KKT) optimality conditions of the first-order reliability method for the solution of two differently formulated RBDO problems, namely the minimization of cost under reliability constraints and the  maximization of reliability under cost constraints. A large number of techniques based on KKT conditions have been proposed since then (See for instance \citet{Kharmanda2002,Agarwal2007, Kaymaz2007}). The most popular mono-level techniques are the \emph{single loop single vector} and \emph{single loop approach}, which are based on approximating the minimum performance target point using the limit-state function sensitivities \citep{Chen1997,Liang2004,Liang2007}. 

Despite the gain in efficiency with respect to two-level approaches, mono-level formulations also suffer from some limitations. In the numerical benchmark carried out by \citet{Aoues2010}, it is shown for instance that SLA and the KKT-based approaches often fail to converge when the starting point of the optimization problem is far from the optimal solution. Lack of robustness of these methods is also observed when target reliability indexes are large or when the design variables are the mean of the random parameters.

\subsubsection{Decoupled approaches} 
As an alternative to mono-level approaches, so-called \emph{decoupled approaches} have been introduced. They consist in solving the RBDO problem through a sequence of deterministic optimization and reliability analysis. In fact an approximate deterministic optimization problem is solved using information from a previous reliability analysis.

Early contributions in this category suggested to approximate the reliability constraints by introducing Taylor series expansions. \citet{Li1994} formulated a linear programming problem where at each iteration, the first-order Taylor series expansion of the reliability index evaluated at the previous cycle is used. \citet{Royset2001} introduced an approach which relies on an semiinfinite optimization algorithm. The most popular decoupled approach is the \emph{sequential optimization and reliability assessment} (SORA) where the probabilistic constraints are translated for deterministic optimization through inverse FORM \citep{Du2004}. The associated deterministic optimization problem is obtained by shifting the random variables using the most probable target point found at the previous reliability cycle. Numerous papers have been developed to improve the efficiency of SORA. For instance, \citet{Cho2011} combine convex linearization to SORA. \citet{Cheng2006} introduced the \emph{sequential approximate programming} (SAP) which relies on KKT. In SAP, a sub-optimization problem with an approximate objective and constraint functions is solved at each iteration. To avoid running the reliability analysis necessary for obtaining the reliability index and its sensitivities at each cycle, the latter quantities are approximated using a recurrence formula. \citet{Zou2006} also suggested a similar approach where the Taylor series are used to expand on the failure probability rather than the reliability index. In all these approaches, sensitivities of the failure probability or of the reliability index are necessary. They may be obtained in some cases using analytical derivations. When such analytical derivations are not available, the authors suggest the use of finite differences, though at a loss of efficiency.

In general, decoupled approaches suffer the same drawback as mono-level approaches. Given that the reliability and optimization problems are expected to converge simultaneously, approximation errors in the early cycles may lead the algorithm in the wrong search direction. This is exacerbated when the initial design is far from the optimal solution and when the limit-state function is highly non-linear.

\subsection{Use of advanced techniques and simulation in RBDO}
The three groups of methods introduced in the previous section rely on approximation methods for computing failure probabilities. The numerous contributions all aimed at reducing the computational burden resulting from a direct solution of the problem. Various applications have shown that their respective efficiency is tightly linked to the problem at hand. To take advantage on the best of each method, some researchers have proposed hybrid formulations which combine methods from the different groups. For instance, \citet{Youn2007,Lim2016,Li2015, Jiang2017} proposed various approaches that can switch from either single-loop to decoupled or double-loop according to some predefined criteria. Even though such approaches may increase the rate of convergence of these approximation-based methods, they remain flawed because they inherit the well-known limitations of FORM, which can cause inaccuracies or even divergence of the optimization when the limit-state function is highly non-linear or in presence of multiple failure regions. As a consequence, other techniques have been developed in an attempt to overcome these limitations. To deal with high curvatures of the limit-state surfaces, one approach has been the introduction of the second-order reliability method (SORM). This was already discussed in \citet{Shetty1998}, while more recently \citet{Stroemberg2017} proposed coupling SORM and sequential quadratic programming (SQP). The MPP-based dimension reduction method \citep{Rahman2004} has also been used to improve the accuracy of failure probability estimates w.r.t. FORM \citep{Rahman2008b,Lee2008b}.

Strategies based on higher-order approximations may attenuate issues due to high-curvature of the limit-state function, however they do not directly tackle the challenges related to the search of the most probable failure or target points (respectively MPP and MPTP). The use of simulation methods in the estimation is in this respect an appropriate solution. Numerous contributions have considered simulation methods in the inner loop of the two-level framework. Due to their high cost, classical simulation techniques are often associated to surrogate models, as discussed in the next section. However, some methods that solely depend on advanced simulation techniques, have been developed. For instance, \citet{Rashki2014} use a weighted simulation method to solve RBDO problems which are however limited to cases with only random design variables (no deterministic and no environmental variables). \citet{Beaurepaire2013} use bridge importance sampling where the idea is to recycle information from previous iterations during the optimization process. Direct integration of the simulation methods to optimization is also possible. This has been done for instance in \citet{Royset2004} where sensitivities of failure probability obtained using Monte Carlo simulation or importance sampling are developed. Finally, \citet{Taflanidis2008} introduced the so-called \emph{stochastic subset optimization} (SSO) for RBDO problems where the failure probability appears in the objective function. Upon formulating an augmented reliability problem where the design variables are artificially considered as uncertain, SSO iteratively identifies subsets of the original design space with high likelihood of containing the optimal design. Identifying sub-regions of the design space is a challenging task which involves a non-smooth optimization problem. To bypass this issue, \citet{Jia2013} proposed to use kernel density estimation to directly approximate the objective function instead of working with subsets. Remaining in the context of augmented reliability problem, \citet{Liu2017} recently proposed a similar method where the failure probability, seen as a function of the design parameters, is approximated.

Another reformulation of the RBDO which is based on quantiles has been introduced by \citet{MoustaphaSMO2016}:
\begin{equation}\label{eq:RBDO_QMC}
\begin{split}
& \ve{d}^\ast = \arg \min_{\ve{d} \in \mathbb{D}} \mathfrak{c} \prt{\ve{d}} \quad \text{subject to: }
\left\{ \begin{array}{ll}
\mathfrak{f}_j\prt{\ve{d}} \leq 0, \quad &  \acc{j = 1 \enum n_s}, \\
Q_{\alpha_k}\prt{\ve{d}; \mathfrak{g}_k\prt{\ve{X}\prt{\ve{d}},\ve{Z}} } \leq 0, \quad & \acc{k = 1 \enum n_h},
\end{array} \right.
\end{split}
\end{equation}
where
\begin{equation}
\label{eq:Q_def}
Q_{\alpha_k}\prt{\ve{d}; \mathfrak{g}_k\prt{\ve{X}\prt{\ve{d}},\ve{Z}}} = \inf\acc{q \in \mathbb{R}: \Prob{\mathfrak{g}_k \prt{\ve{X}\prt{\ve{d}},\ve{Z}} \leq q} \geq \alpha_k },
\end{equation}
and $\alpha_k = \bar{P}_{f_k}$. Such a formulation has been used in the literature under the name of \emph{percentile approach} (as used in PMA and SORA) whose solution relies on the so-called minimum target performance point (MPTP), located using inverse FORM. Eq.~(\ref{eq:RBDO_QMC}) solves the same problem but in a more direct manner where Monte Carlo simulation is considered for the computation of the quantile. The latter can be seen as an equivalent of the limit-state function evaluated at the MPTP.  

All these approaches based on advanced simulation techniques have brought a substantial gain in model evaluation savings, \ie going from $10^8-10^9$ for a direct two-level approach with crude Monte Carlo simulation to $10^4-10^5$ calls to the performance function. However, they still remain expensive when considering time-consuming performance function evaluations. For practical applications, efficient methodologies associate simulation methods with surrogate modelling.

\subsection{Use of surrogate models in RBDO}
\begin{figure}[!ht]
\begin{center}
\includegraphics[width=0.49\textwidth]{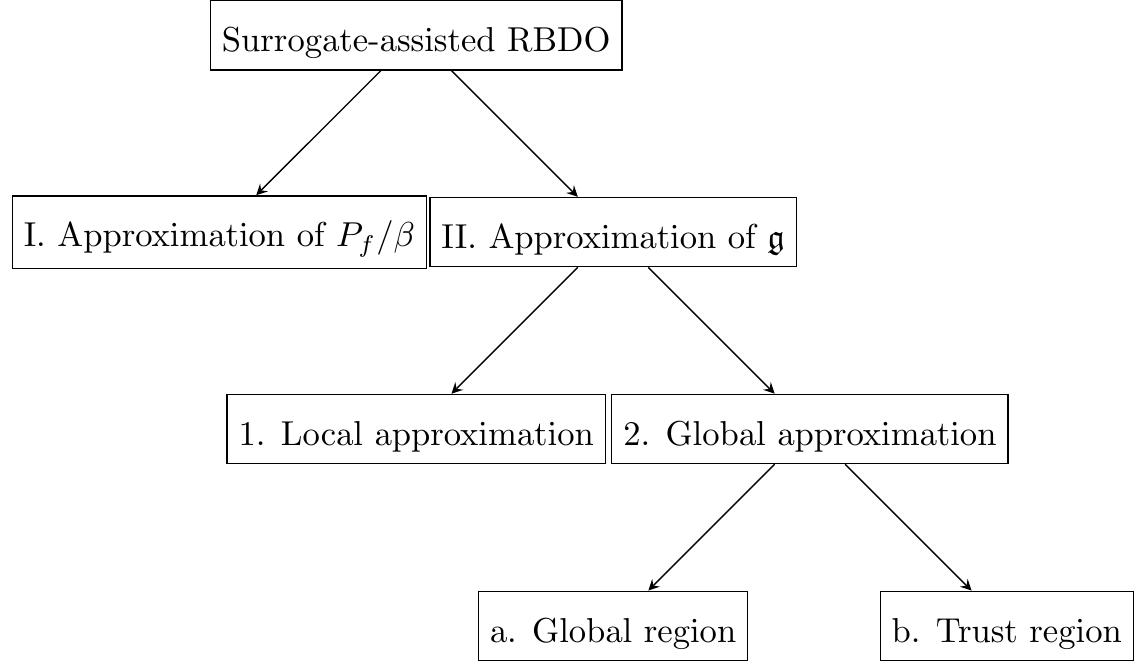}
\end{center}
\caption{Flowchart of different surrogate-assisted RBDO schemes}
\label{fig:MetaRBDO}
\end{figure}

The basic idea in surrogate modelling is to replace an expensive-to-evaluate function with a cheap approximation constructed using an \emph{experimental design}, a.k.a \emph{training set}, consisting of a limited number of evaluations of the original model. Various surrogate models have been employed in RBDO problems using different solution techniques, \ie based on simulation or approximation methods. However, the schemes in which they are integrated differ from one contribution to the other. This is summarized in the flowchart of \figref{fig:MetaRBDO}.

The first difference that exists between the various approaches in the literature relates to the output function that is actually surrogated. One particular approach consists in directly approximating the relationship between a given design and the corresponding failure probability (resp. reliability index) (branch I. in \figref{fig:MetaRBDO}). In other words, the experimental design consists of $N$ pairs $\acc{\prt{\ve{d}^{(i)}, P_f\prt{\ve{d}^{(i)}}}, i = 1, \ldots, N}$ (resp. $\acc{\prt{\ve{d}^{(i)}, \beta\prt{\ve{d}^{(i)}}}, i = 1, \ldots, N}$ ). The optimization can then proceed using any suitable algorithm and replacing the true failure probability with the one estimated by the surrogate model. This approach has been used in various recent contributions. For instance, \citet{Foschi2002} use response surface modelling to approximate the relationship between different design choices and the corresponding reliability indexes computed by importance sampling. More recently, \citet{Lekhy2017} developed an approach where the failure probability computed through FORM is approximated using artificial neural networks (a type of surrogate model in itself). In these approaches, one sample of the training set requires a full reliability analysis based on the true model. This becomes problematic when the size of the experimental design is large and/or when the performance functions are expensive-to-evaluate. 

A less computationally expensive alternative consists in directly creating an approximation of the performance function and to use it for reliability analysis (branch II. in \figref{fig:MetaRBDO}). Again, building on this principle, various schemes have been proposed recently. One obvious approach is to build distinct surrogate models locally used for each reliability analysis in the inner loop of a two-level approach (branch II.1 in \figref{fig:MetaRBDO}). \citet{Agarwal2004} carry out a double-loop approach where at each iteration, a second-order response surface model is built around the MPP of the previous iteration in order to perform an approximate reliability analysis. Similarly, \citet{Papadrakakis2005} locally trains a neural network and use it to compute a Monte Carlo estimate of the failure probability at each iteration of a double-loop optimization process. This approach is also not optimal as it requires building from scratch a surrogate model at each iteration of the optimization process. 

Another option that has been explored is building a single surrogate model that can be used to assess the failure probabilities considering multiple design choices (branch II.2 in \figref{fig:MetaRBDO}). The challenge with such an approach is to build an accurate surrogate model over a large area. An important observation though is that there is no need for a surrogate model that is highly accurate globally, as only sub-regions will be explored during optimization. As a consequence, most of the methods developed in this category resort to \emph{adaptive sampling} in order to ensure accuracy only in regions of interest. Various methods have been developed depending on the space chosen to build the surrogate model. In problems where only design variables are considered to be possibly uncertain, the space over which the surrogate model is built may coincide with the design space (branch II.2.a in \figref{fig:MetaRBDO}). In \citet{Chen2015b}, a Kriging model is built over the whole design space and SORA is used as the RBDO method. \citet{Li2016} solve a RBDO problem using a Kriging model defined on the design space together with importance sampling. In the two contributions, the surrogate models are built adaptively using respectively the importance boundary sampling and a local sampling method based on the MPP. 

Other authors have considered to rather build a surrogate model in a sub-region of the space rather than a global one (branch II.2.b in \figref{fig:MetaRBDO}). This sub-region, often referred to as \emph{local window} or \emph{trust region}, consists of a hypercube in a \emph{hybrid space} encompassing design and random variables \citep{Kharmanda2002} whose size and center is updated as the optimization progresses. The difference between the local approximation (branch II.1) and the global approximation with trust region (branch II.2.b) is that in the former, the built surrogate model can only be used with the current value of the design parameters whereas in the latter, the input space allows for the computation of the failure probability for different designs using the same surrogate model. As an example of methods based on trust-region, \citet{Lee} builds dynamic Kriging models on local windows and associates this with MCS for the solution of RBDO problems. Other similar contributions include \citet{Taflanidis2014,Zhang2017,Gao2017,Gaspar2017}. \citet{Lee2011,SongThesis2013} argue that the accuracy of the surrogate model can be increased, specially for high-dimensional problems, if the local window is hyperspherical. They then propose methods respectively based on virtual support vector machines or dynamic Kriging.

In most of the contributions listed above, it is not clear how the design space and the reliability space are explored simultaneously by the surrogate model. A rigorous framework has been established based on the idea of \emph{augmented reliability space} \citep{Au2005}. The associated scheme can be classified in the group labeled II.2.a in \figref{fig:MetaRBDO}. The original principle is to artificially consider design variables as uncertain and then use conditional sampling and Bayes theorem to compute the failure probability given a realization of the design variables. Adopting this philosophy, \citet{TaflanidisThesis2007, Taflanidis2008} formulate a problem where the failure probability lies in the objective function and then construct the stochastic subset optimization algorithm. \citet{Taflanidis2014,Zhang2017} solve a similar problem, now using adaptive Kriging and trust regions. \citet{Dubourg2011} also developed the so-called \emph{meta-IS} approach by combining augmented space, adaptive Kriging and importance sampling. Finally, \citet{MoustaphaSMO2016,MoustaphaICOSSAR2017} formulate a quantile-based RBDO problem and solve it using Monte Carlo sampling based respectively on Kriging and polynomial chaos expansions models built over an augmented space. 

The use of an augmented space allows one to rigorously solve problems where all combination of deterministic/random and design/environmental variables can be considered. Starting from this premise, an original generalized framework for RBDO combining adaptive surrogate modelling and simulation is now proposed.

\section{Proposed framework}
\subsection{Motivation}

As shown in the previous review section, there is a tremendous amount of methods developed for the solution of RBDO problems. Early methods were based on local approximations whose aim was to reduce the associated computational burden. Even though these approaches can solve a fair amount of problems, they are limited by their lack of robustness due to the very use of FORM. Recent papers attempted to face these limitations but their computational cost remain relatively large, thus limiting their range of applications to simple toy problems. As a consequence, methods based on simulation methods have been developed to bypass the FORM-related issues. The improvement of accuracy and robustness brought by these methods however comes with of a dramatic increase of the computational cost. They have therefore been associated to surrogate modelling to solve real-world problems where the performance function relies on expensive-to-evaluate models.

Various methodologies that combine surrogate modelling and advanced reliability techniques to solve RBDO problems have been proposed. However as indicated in the previous section, these methods are either able to solve only a limited class of problems or are problem-specific \emph{i.e.}, their implementations require advanced knowledge and specialized derivations. In this paper, we propose a generalized framework based on simulation and adaptive surrogate modelling for the solution of RBDO problems. The interest of the proposed framework can be delineated in the following three points. 

First, the framework is based on a formulation where all possible combinations of random/deterministic on the one hand and design/environmental variables on the other hand, can be accounted for at once. Second, this modular framework is \emph{non-intrusive}, \ie it consists of three distinct and \emph{independent} blocks, namely surrogate modelling, structural reliability and optimization. There is no coupling between the different blocks in the basic form of the proposed framework. For instance, the fact that we do not consider the use of analytical gradients for optimization allows us to avoid specific implementations related to the sensitivity of failure probability (or reliability index), which would depend on the reliability technique used. Finally, there is no specific requirement as to which method goes to each block. This will be illustrated in the next section where we will consider applications with different techniques for each of the three blocks namely \ie Kriging \vs support vector regression, Monte Carlo simulation \vs subset sampling and gradient-based \vs gradient-free optimization algorithms.
\begin{figure}[!ht]
\centering
\includegraphics[width=1\textwidth]{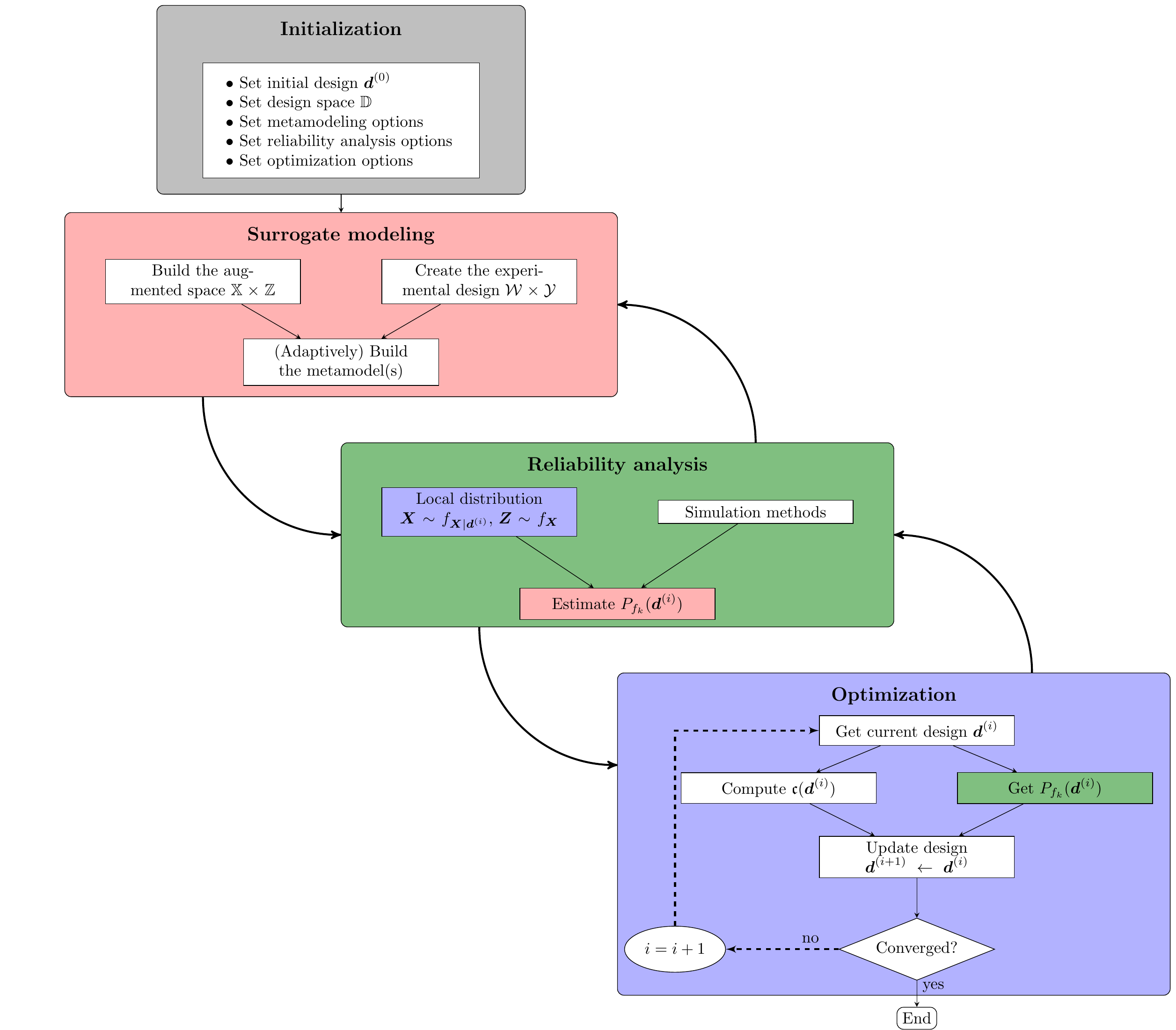}
\caption{Flowchart of the proposed framework.}
\label{fig:framework}
\end{figure}

\subsection{Detailed implementation of the approach}
As shown in \figref{fig:framework}, beside initialization, the proposed framework consists of three different blocks which are now detailed.

\subsubsection{Surrogate modelling}\label{sec:surrogate modelling}
\paragraph{Basic idea\\}
Surrogate modelling is in the core of the proposed framework. The basic idea is to replace any expensive-to-evaluate black-box function by an easy-to-evaluate analytical function. This is made possible by assuming that the original model has some accommodating properties such as regularity/continuity and follows a general functional shape. Based on different assumptions, many different surrogate modelling techniques are nowadays available, among which Gaussian process modelling (a.k.a Kriging), polynomial chaos expansions, support vector machines, artificial neural networks, etc. Each of these surrogate models possesses some structure and parameters that need to be calibrated through a learning algorithm on a limited set of data also known as the experimental design or training set: $\mathcal{D} = \acc{(\mathcal{W}^{(i)}, \mathcal{Y}^{(i)}), i= 1,\ldots, N}$, where $\mathcal{W}^{(i)}$ is an input of the computational model $\mathcal{M}$ and $\mathcal{Y}^{(i)} = \mathcal{M}(\mathcal{W}^{(i)})$ is the associated response.

\paragraph{Augmented space \\} 
As explained in the previous section, there are various ways of building a surrogate model for RBDO. In this work, we consider building a single surrogate model in an augmented space which spans both the space of random or deterministic design parameters and environmental variables. In most cases, the corresponding input space is a bounded hypercube. The augmented space considered here aims at defining a confidence region over which the surrogate model will be evaluated so as to minimize extrapolation, \ie evaluation of the surrogate model far from existing training points. More precisely, it consists of a tensor product of confidence regions defined for the design parameters and environmental variables. For the former, we proceed by simply extending the deterministic bounds on the design parameters $d_i$ in each dimension to reduce to an acceptable level the probability of sampling outside this area for designs close to bounds. Let us denote these probabilities by $\alpha_{d_i^-}$ and $\alpha_{d_i^+}$ respectively for the lower and upper bounds in each dimension $i = \acc{1, \ldots, M_d}$. The bounds of the associated augmented space in the $i$-th dimension therefore read:
\begin{equation}
x_i^- = F^{-1}_{X_i|d_i^-}(\alpha_{d_i^-}), \qquad x_i^+ = F^{-1}_{X_i|d_i^+}(1 - \alpha_{d_i^+}),
\end{equation}
where $F^{-1}_{X_i|d_i^-}$ and $F^{-1}_{X_i|d_i^+}$ are respectively the inverse cumulative distribution functions (CDF) of the random variables associated to lower and upper bounds of $d_i$, denoted by $d_i^-$ and $d_i^+$.
The design confidence region is eventually obtained by the following tensorization:
\begin{equation}
\mathbb{X} = \prod_{i=1}^{M_d} \bra{x_i^-,x_i^+}.
\end{equation}
For an environmental variable, the marginal confidence region is obtained by defining a symmetric bounded area (with respect to the probability content) around its mean value. By defining the above probability by $\alpha_z$, the marginal confidence region can be obtained as follows:
\begin{equation}\label{eq:z-z+}
z_i^- = F^{-1}_{Z_i}(\alpha_{z_i}/2), \qquad z_i^+ = F^{-1}_{Z_i}(1 - \alpha_{z_i}/2),
\end{equation}
where $F^{-1}_{Z_i}$ is the inverse CDF associated to the environmental variable $Z_i$.
The associated confidence region therefore reads:
\begin{equation}\label{eq:marg:Z}
\mathbb{Z} = \prod_{i=1}^{M_z} \bra{z_i^-,z_i^+}.
\end{equation}

Note that for the environmental variables, the construction of $\mathbb{Z}$ is not limited to this hypercube. In fact, since the environmental variables do not evolve during the optimization, the original support of these random variables can be directly used as marginal confidence region $\mathbb{Z}$. Also, transformed variables $\ve{U} = \mathbb{T}\prt{\ve{Z}}$ in the usual standard normal space could be considered. In this paper, we however consider the construction in \eqref{eq:marg:Z}, which to the authors experience leads to more accurate surrogate models.

Finally, the augmented space follows as a tensor product of the defined marginal confidence regions, \emph{i.e.}:
\begin{equation}
\mathbb{W} = \mathbb{X} \times \mathbb{Z}.
\end{equation}

\begin{figure}[!ht]
\centering
\includegraphics[width=0.75\textwidth]{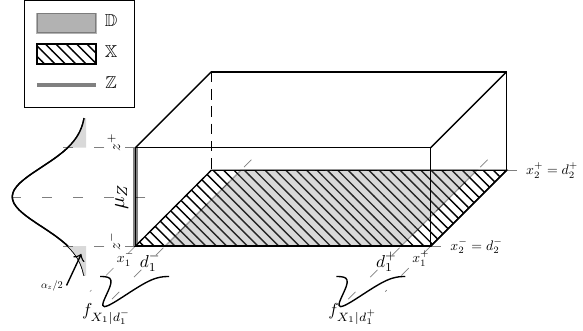}
\caption{Illustration of the augmented space for a three-dimensional problem considering two design variables $X_1\prt{d_1}$ and $d_2$ and a random environmental variable $Z$.}
\label{fig:AugSpa}
\end{figure}

An illustration of an augmented space for a three-dimensional problem is shown in \figref{fig:AugSpa}. This problem consists of two design- and one environmental variables. Design parameter $d_1$ is the mean value of random variable $X_1$, while design parameter $d_2$ is deterministic. This is directly accounted for in the construction of $\mathbb{X}$ by extending the bounds in the first dimension only, \ie $x_2^-=d_2^-$ and $x_2^+ = d_2^+$. The flexibility of this construction allows one to consider all possible combinations of deterministic or random variables and the presence or not of random environmental variables. 

\paragraph{Active learning \\} 
Once the augmented space is defined, the next step is to sample points that will be evaluated using the original model in order to generate a training set. A common approach is to sample the space uniformly. This can be achieved using stratified sampling approaches such as the well-known \emph{Latin hypercube sampling} \citep{McKay1979} or low-discrepancy sequences such as the \emph{Sobol'} sequence \citep{Sobol1967}. To reach the level of surrogate model accuracy necessary for a valid RBDO solution, the space needs to be densely sampled. Unfortunately, in some cases the resulting training set would be so large that it would defy the purpose of using surrogates in the first place. However, it can be observed that the accuracy of the surrogate model does not necessarily need to be large over the entire space but only in some limited regions of interest. This observation gave rise to approaches known as \emph{adaptive design} or \emph{active learning}, where the idea is to start with a small training set and then gradually increase its size by strategically adding points to improve the accuracy of the surrogate model following the requirements of the analysis at hand (reliability analysis or optimization). 

When a limit-state function is approximated, the region of interest lies in the vicinity of the limit-state surface. Various enrichment methods exist in this perspective. By various means, they all aim to \emph{explore} the space \ie adding points in areas which are scarce in data and \emph{exploit} the space by focusing on the vicinity of the approximated limit-state surface.

Active learning is used in this framework to reduce the number of calls to the original model. This allows one to use a \emph{single} global surrogate model for the reliability analyses in each iteration of the optimization process.

\subsection{Structural reliability analysis}
The structural reliability analysis block is considered as a black-box module in the proposed framework. This means that any method to estimate failure probabilities that is judged most suitable for the problem at hand can be used. In this work, focus is given to simulation methods rather than approximation techniques due to the fact that the limit-state surface is an inexpensive-to-evaluate surrogate model. The former are known to yield more accurate results, at the expense of a dramatically increased computational cost though.  Naturally, crude Monte Carlo simulation is the most robust approach when it comes to estimating failure probabilities. However the variance of the estimates decays very slowly with respect to the number of samples. This becomes even more problematic in the presence of a rare event (\ie with an extremely low failure probability). Advanced simulation methods  such as subset simulation \citep{Au2001, Papaioannou2015}, cross-entropy importance sampling \citep{Kurtz2013, Geyer2019} or line sampling \citep{Pradlwarter2007,DeAngelis2015} that try to achieve a faster decrease of the variance of $P_f$ with diminishing values of the target failure probability have been developed and shall be used whenever justified.
 
\subsection{Design optimization}
\paragraph{Global \vs local \\}
As stated earlier, any optimization algorithm can be plugged non-intrusively in the proposed framework. In general, one must distinguish so-called \emph{global} from \emph{local} search methods. In the former, the entire design space is explored by sampling points according to a mechanism proper to the algorithm. This sampling procedure is expected to converge towards the region with the highest likelihood of containing the optimal solution. In contrast, the latter starts with an initial guess and generates a sequence of improved designs using some local information on the cost function. The sequence is expected to converge to a local minimizer which cannot be guaranteed to be the global one, except if some conditions are met, \eg{} the function is convex. In contrast, global algorithms possess internal mechanisms that increase the likelihood of reaching the global optimizer, at usually a typical higher cost w.r.t. local optimizers.

\paragraph{Variance reduction\\}
A peculiar aspect of the RBDO problem is that the constraints are stochastic in nature. In such a context, the computation of the sensitivities required by gradient-based approaches is not trivial. For some reliability methods, it is possible to derive the sensitivities of the failure probability with respect to the design variable, for instance using the so-called \emph{score functions} together with crude Monte Carlo simulation. Sensitivities of the cost with respect to the design variables through analytical functions can improve the efficiency of the optimization algorithm. However, such information may not be available and in any case it is not general, \ie it depends on the reliability method or distribution functions of the random variables. This defies the purpose of providing a generalized non-intrusive framework. 

In this work, we rather suggest the numerical estimation of the sensitivities. One widespread approach relies on \emph{finite-differences} where the gradient can be estimated at the cost of $M_d+1$ (forward or backward) or $2M_d$ (central finite difference) evaluations of the original model. One of the issues with this approach is the stochasticity in the failure probability. In fact, the noise inherent to each estimation of $P_f$ or $\beta$ in the finite difference procedure (due to the finite sets of Monte Carlo sample) could make it difficult to properly estimate the gradient, as one cannot anymore differentiate the information given by the actual perturbation of the design from that due to random noise. To avoid this issue, it is possible to resort to the so-called \emph{common random numbers} approach \citep{Spall2003_ch14}. This consists in introducing a consistent error between the estimates of the failure probabilities for two different designs. In practice, this may be achieved by using the same stream of random numbers to generate the samples needed to estimate $P_f$. \emph{Exterior sampling approach} (ESA) is similar to CRN as the same stream of random numbers is used throughout the entire optimization process \citep{Taflanidis2008}. This actually transforms the stochastic optimization problem into a deterministic one that can be solved using any general-purpose optimization algorithm. However, the solution may be biased due to the introduction of the consistent error. This can be accounted for by either increasing the sample size or repeating the optimization procedure multiple times. 

\paragraph{Stochastic approximation methods\\}
The use of common random numbers or ESA however, does not solve all challenges related to RBDO using finite-difference-based approaches. For instance, the finite difference step size should be consistent with the variance of the estimated failure probability. Besides, for some advanced simulation methods such as subset simulation or importance sampling, the use of the ESA does not necessarily translate into a deterministic optimization problem. In fact the topology in the reliability analysis can change drastically between two infinitesimally close designs. An alternative approach to approximate gradients is based on \emph{simultaneous perturbation} \citep{Spall1998a,Spall1998b} where, instead of varying one component of the design at a time (as in finite-difference), one can simultaneously and randomly perturb all components of the design. The simultaneous perturbation stochastic approach (SPSA) \citep{Spall2003} is based on this principle and has the advantage of requiring only two function evaluations for each gradient estimate, regardless of the problem dimension. \citet{Taflanidis2008} explore its use in RBDO where the formulation consists solely in minimizing the failure probability. This does not directly apply to the formulation considered here as the stochastic part is in the constraints. One may however consider constrained formulations of SPSA as proposed for instance in \citet{Wang1998,Wang2003}.

In our framework, we focus more on global stochastic algorithms. The reasons are two-fold. First, as discussed earlier, such methods have a greater likelihood to find the global optimizer. The associated increase in computational effort can be leveraged by the fact that the cost function is a simple analytical function and that the failure probability estimate is based on a surrogate model. Second, the fact that only first-order information is needed allows one to avoid issues related to computing sensitivities of the failure probability which can be, at best, inefficient.
%%%%%%%%%%%%%%%%%%%%%%%%%%%%%%%%%%%%%%%%%%%%%%%%%%%%%%%%%%%%%%%%%%%%%%%%%%%%%
\section{Case studies}
In this section, we will consider three different configurations of the proposed framework to solve a set of benchmark problems. For each configuration case, we will consider different methods in each block as detailed in Table~\ref{tab:01}. The reference solution corresponds to actually using the original model associated to a large Monte Carlo set for the reliability analysis. The optimization algorithm is hybrid covariance matrix adaptation - evolution scheme (CMA-ES), meaning that we start with the CMA-ES \citep{Hansen2001,Arnold2012} to locate the global minimum and refine the found solution using a gradient-based algorithm. The entire analysis is carried out in \textsc{UQLab}, a \textsc{Matlab}-based framework for uncertainty quantification \citep{MarelliUQLab2014}. The classical algorithms presented in the review are available in the RBDO module of \textsc{UQLab} together with the proposed algorithm \citep{UQdoc_12_114}.
\begin{table}[h]
	\centering	
	\begin{tabular}{cccc}
		\hline
		Framework & Surrogate model & Reliability method & Optimization algorithm \\ \hline
		Case $\#1$    &  SVR   &  Monte Carlo simulation  &           SQP           \\
		Case $\#2$    &  Kriging   & Subset simulation &     CMA-ES           \\
		Case $\#3$     &  Kriging   & Quantile Monte Carlo  &		CMA-ES with enrichment            \\
		Reference     &  Original model   & Monte Carlo simulation  &           Hybrid CMA-ES           \\ \hline
	\end{tabular}
	\caption{Different realizations of the proposed framework to be used in the benchmark problems.}
	\label{tab:01}
\end{table}

\subsection{Case $\# 1$}

In the first case, we consider adaptive support vector regression as a surrogate model, crude Monte Carlo simulation for the reliability analysis and the sequential quadratic programming (SQP) as provided by \textsc{Matlab}'s \texttt{fmincon} as the gradient-based optimization algorithm. As the latter two methods are standard, we focus only on support vector regression and its adaptive enrichment in the sequel. 

\paragraph{Basics of SVR\\}
\noindent
Support vector regression \citep{Vapnik:1995,Smola2004,UQdoc_11_111} is a machine learning technique often used for surrogate modelling in structural reliability. Given a training set $\mathcal{D} = \acc{\mathcal{W},\mathcal{Y}}$ as defined in Section~\ref{sec:surrogate modelling}, the SVR prediction for a new point reads:
\begin{equation}\label{eq:svr}
\mathcal{M}^{\text{SVR}}\prt{\ve{w}} = \sum_{i=1}^N \prt{\alpha_i - \alpha_i^\ast}  k \prt{\boldsymbol{w}_i, \boldsymbol{w}} + b,
\end{equation}
where $\ve{\alpha}^{(\ast)} = \acc{(\alpha_i, \alpha_i^\ast), i= 1, \ldots, N}$ is a set of unknown coefficients, $b$ is an unknown bias term and $k\prt{\bullet, \bullet}$ is a predefined kernel function. The unknown parameters are found by minimizing a regularized loss function. In this paper, we consider the quadratic $\epsilon$-insensitive loss function defined by $\ell(y, \mathcal{M}^{\text{SVR}}\prt{\ve{w}}) = \prt{\abs{y -  \mathcal{M}^{\text{SVR}}\prt{\ve{w}} } - \varepsilon}^2$ if $\abs{y -  \mathcal{M}^{\text{SVR}}\prt{\ve{w}} } < \varepsilon$ and $0$ otherwise. In practice, the coefficients $\ve{\alpha}^{(\ast)}$ correspond to Lagrange multipliers that can be found by minimizing the following functional \citep{BourinetHDR}:
\begin{equation}\label{eq:QOP}
\begin{split}
\arg \min_{\ve{\alpha},\ve{\alpha}^\ast} & \qquad \frac{1}{2}\prt{\ve{\alpha} - \ve{\alpha}^\ast}\widetilde{\ve{K}}\prt{\ve{\alpha} - \ve{\alpha}^\ast}  + \sum_{i=1}^{N} \prt{\alpha_i + \alpha_i^\ast} \varepsilon + \sum_{i=1}^{N} \prt{\alpha_i - \alpha_i^\ast} y_i, \\
\text{subject to:} &  \qquad \sum_{i=1}^{N}  \prt{\alpha_i - \alpha_i^\ast} = 0, \quad \text{and} \quad \alpha_i, \alpha_i^\ast \geq 0, \; i= 1,\ldots, N,
\end{split}
\end{equation}
where $\widetilde{\ve{K}} = \ve{K} + 1/C \ve{I}_N$ with $\ve{K} = \bra{k\prt{\ve{w}_i,\ve{w}_j}}_{1\leq i,j \leq N}$ and $\ve{I}_N$ being respectively the Gram and identity matrices of size $N\times N$.

Eq.~\eqref{eq:QOP} is a convex quadratic optimization programming problem which can be solved using any specialized solver. The remaining bias term $b$ in Eq.~\eqref{eq:svr} can be obtained as by-product of the solution of this problem. 

In general, one needs to calibrate the model \ie find the optimal hyper-parameters, namely the penalty term $C$, the $\varepsilon$-insensitive tube width $\varepsilon$ and the parameter(s) of the chosen kernel function. This is done here by minimizing an approximation of the leave-one-out (LOO) cross-validation error, more specifically the so-called \emph{span estimate} \citep{Chapelle2002}. The LOO error is aimed at approximating the surrogate model generalization error. The overall optimization problem is carried out here using the covariance matrix adaptation-evolution scheme (CMA-ES) (\citet{Hansen2001}, see also \citet{BourinetHDR,MoustaphaJRUES2018}).

\paragraph{Adaptive construction of the training set\\}
Various methods have been proposed for the adaptive construction of support vector machines in the context of structural reliability. We consider a modified version of the method proposed in \citet{Basudhar2010} where the authors suggest to solve the following maximin problem:
\begin{equation}\label{eq:maxmin}
\max_{\ve{w} \in \mathbb{W}} \min_{i=1,\ldots,N} \lVert \ve{w}  - \ve{w}^{(i)} \rVert  \quad\text{subject to:} \quad \mathfrak{g}\prt{\ve{w}} = 0.
\end{equation}
In other words, this problem consists in finding the point from the currently approximated limit-state surface (exploitation) that is the furthest away from the existing training set (exploration).

Finding the next point to add however consists in solving a continuous maximin problem which may be expensive. In this paper, we propose to discretize this formulation. In this respect, we first sample a set of candidates for enrichment $\mathcal{S} = \acc{\ve{s}^{(1)}, \enum \ve{s}^{(N_s)}}$. Then a subset $\mathcal{S}^\prime$ of this set consisting of the closest points to the limit-state surface is selected. Finally the next best point is chosen by adapting Eq.~\eqref{eq:maxmin}, \ie:\\
\begin{equation}\label{eq:maxmin_dsic}
\ve{s}^{\text{next}} =	\max_{\ve{s} \in \mathcal{S}^\prime} \min_{i=1,\ldots,N} \lVert \ve{s}  - \ve{w}^{(i)} \rVert,
\end{equation}
where $\mathcal{S}^\prime = \acc{\ve{s} \in \mathcal{S}: \lvert \widehat{\mathfrak{g}}\prt{\ve{s}} \rvert \leq  q}$ with $q$ being an $\gamma$-quantile of $\lvert \widehat{\mathfrak{g}}\prt{\ve{s}} \rvert$. In this paper, we set $\gamma = 0.01$.

Two convergence criteria are considered here. The first first one is related to the stability of the failure domain within iterations of enrichment and reads:
\begin{equation}\label{eq:enr_conv1}
\frac{\lvert N_f^{(i-1)} - N_f^{(i)} \rvert}{N_f^{(i-1)} } < \epsilon_{D_f}
\end{equation}
where $N_f^{(i)}$ is the number of failed samples in $\mathcal{S}$ as computed at the $i$-th enrichment iteration.

The second criterion relates to the stability of failed \vs safe predictions: the surrogate model is assumed to be good enough when the number of points that are predicted in the current safe domain while it belonged to the failure domain in the previous iteration (and vice-versa) is low. This can be written as:
\begin{equation}\label{eq:enr_conv2}
\frac{\text{Card} \prt{\ve{s} \in \mathcal{S}: \widehat{\mathfrak{g}}^{(i-1)}\prt{\ve{s}} \times \widehat{\mathfrak{g}}^{(i)}\prt{\ve{s}} < 0 } }{N_s} < \epsilon_{SC},
\end{equation}
where  $\widehat{\mathfrak{g}}^{(i)}$ denotes the surrogate model at the $i$th iteration.

Convergence is assumed in the following applications when these two criteria are satisfied in two iterations in a row. The thresholds are set to $\epsilon_{D_f} = 0.001$ and $\epsilon_{SC} = 0.001$ in the case studies presented in Section~\ref{sec:Applications}.
\subsection{Case $\#2$}\label{sec:AK}

\paragraph{Basics of Kriging\\}
Kriging \citep{Santner2003, Rasmussen2006,Lataniotis2018} is a surrogate modelling technique which considers the model $\mathcal{M}$ as a realization of a stochastic Gaussian process that can be cast as:
\begin{equation}
\mathcal{M}\prt{\ve{w}} = \sum_{j=1}^p \beta_j f_j\prt{\ve{w}} + Z\prt{\ve{w}},
\end{equation}
where the so-called \emph{trend} consists of a linear combination of $p$ preselected basis functions $\ve{f} = \acc{f_j, j = 1, \ldots, P}$ and $Z$ is a zero-mean stationary Gaussian process with auto-covariance $\text{Cov}\bra{Z\prt{\ve{w}},Z\prt{\ve{w}'}} = \sigma^2 R\prt{\ve{w},\ve{w}'; \ve{\theta}}$. In the last equation, $\sigma^2$ is a constant variance of the process and $R$ is a preselected auto-correlation function with parameters $\ve{\theta}$.

The calibration of this model consists in estimating the regression coefficients \newline $\ve{\beta} = \acc{\beta_j \; j = 1 \enum p}$ and the parameters $\ve{\theta}$ of the selected auto-correlation function. The weight coefficients can be estimated through least-square regression while the auto-correlation parameters can be estimated through cross-validation and maximum likelihood estimation \citep{Bachoc2013b}. Once the calibration is done, the prediction for a new point $\ve{w}$ is assumed to follow a Gaussian distribution \ie{} $\mathcal{M}^{\text{K}}\prt{\ve{w}} \sim \mathcal{N} \prt{ \mu_{\mathcal{M}^{\text{K}}}\prt{\ve{w}}, \sigma^2_{\mathcal{M}^{\text{K}}} \prt{\ve{w}} }$, with
\begin{equation}\label{eq:KRG_MS}
\begin{split}
&  \mu_{\mathcal{M}^{\text{K}}}\prt{\ve{w}} = \ve{f}^T \prt{\ve{w}} \widehat{\ve{\beta}} + \ve{r}^T\prt{\ve{w}} \mat{R}^{-1} \prt{\ve{y} - \mat{F}^T \widehat{\ve{\beta}}},\\
& \sigma^2_{\mathcal{M}^{\text{K}}} \prt{\ve{w}} = \sigma^2 \prt{1 - \ve{r}^T\prt{\ve{w}} \mat{R}^{-1} \ve{r}\prt{\ve{w}} + \ve{u}^T\prt{\ve{w}} \prt{\mat{F}^T \mat{R}^{-1}\mat{F}}^{-1} \ve{u}\prt{\ve{w}}},
\end{split}
\end{equation}
where $\widehat{\ve{\beta}} = \prt{\mat{F}^T \mat{R}^{-1} \mat{F}}^{-1} \mat{F}^T \mat{R}^{-1} \ve{y}$ is the generalized least-square estimate of the weight coefficients $\ve{\beta}$, $\ve{r}\prt{\ve{w}}$ is a vector of cross-correlations between the point $\ve{w}$ and each point of the training set with components $r_j = k \prt{\ve{w}, w_j}, j = 1, \enum, N$. The following notations have been introduced for the sake of clarity:  $\mat{F} = \bra{f_j\prt{\ve{w}_i}}_{1\leq i \leq N, \, 1 \leq j \leq p}$  and $\ve{u}\prt{\ve{w}} = \mat{F}^T \mat{R}^{-1} \ve{r}\prt{\ve{w}} - \ve{f}\prt{\ve{w}}$.

\paragraph{Adaptive construction of the training set\\}
In this contribution, the learning function introduced by \citet{Bichon2008} is considered. This consists in choosing the point which maximizes the so-called \emph{expected feasibility function} as the next point to add in the experimental design:
\begin{equation}\label{eq:EFF_dsic}
\ve{s}^{\text{next}} =	\max_{\ve{s} \in \mathcal{S}} EFF \prt{\ve{s}},
\end{equation}
where
\begin{equation} \label{eq:EFF}
\begin{split}
EFF(\ve{s})  = \mu_{\mathcal{M}^{\text{K}}}\prt{\ve{s}}  \Bigg[ 2 \Phi  \bigg( \frac{\mu_{\mathcal{M}^{\text{K}}}\prt{\ve{s}}}{\sigma_{\mathcal{M}^{\text{K}}}\prt{\ve{s}}} \bigg) - \Phi  \bigg( \frac{-2 \sigma_{\mathcal{M}^{\text{K}}}\prt{\ve{s}} -\mu_{\mathcal{M}^{\text{K}}}\prt{\ve{s}}}{\sigma_{\mathcal{M}^{\text{K}}}\prt{\ve{s}}} \bigg) - \Phi  \bigg( \frac{2 \sigma_{\mathcal{M}^{\text{K}}}\prt{\ve{s}} -\mu_{\mathcal{M}^{\text{K}}}\prt{\ve{s}}}{\sigma_{\mathcal{M}^{\text{K}}}\prt{\ve{s}}} \bigg) \Bigg] \\  - \sigma_{\mathcal{M}^{\text{K}}}\prt{\ve{s}} \Bigg[  2 \varphi  \bigg( \frac{\mu_{\mathcal{M}^{\text{K}}}\prt{\ve{s}}}{\sigma_{\mathcal{M}^{\text{K}}}\prt{\ve{s}}} \bigg) - \varphi  \bigg( \frac{-2 \sigma_{\mathcal{M}^{\text{K}}}\prt{\ve{s}} -\mu_{\mathcal{M}^{\text{K}}}\prt{\ve{s}}}{\sigma_{\mathcal{M}^{\text{K}}}\prt{\ve{s}}} \bigg) - \varphi  \bigg( \frac{2 \sigma_{\mathcal{M}^{\text{K}}}\prt{\ve{s}} -\mu_{\mathcal{M}^{\text{K}}}\prt{\ve{s}}}{\sigma_{\mathcal{M}^{\text{K}}}\prt{\ve{s}}} \bigg) \Bigg]   \\  + 2 \sigma_{\mathcal{M}^{\text{K}}}\prt{\ve{s}} \Bigg[ \Phi \bigg( \frac{2 \sigma_{\mathcal{M}^{\text{K}}}\prt{\ve{s}} - \mu_{\mathcal{M}^{\text{K}}}\prt{\ve{s}}}{\sigma_{\mathcal{M}^{\text{K}}}\prt{\ve{s}}}\bigg) -  \Phi \bigg( \frac{-2 \sigma_{\mathcal{M}^{\text{K}}}\prt{\ve{s}} - \mu_{\mathcal{M}^{\text{K}}}\prt{\ve{s}}}{\sigma_{\mathcal{M}^{\text{K}}}\prt{\ve{s}}}\bigg) \Bigg].
\end{split}
\end{equation}

Enrichment is made sequentially and the convergence criteria in Eqs.~\eqref{eq:enr_conv1} and \eqref{eq:enr_conv2} introduced for SVR are used here as well.
\subsection{Case $\#3$}
In the case $\#3$, we consider the quantile-based formulation introduced in Eq.~\eqref{eq:RBDO_QMC}. Furthermore, a slight modification in the enrichment scheme is introduced in order to increase the performance compared to case $\#2$. In fact, the enrichment procedure introduced above seeks to produce a surrogate model that is accurate in the vicinity of the limit-state surface in the entire augmented space. However optimization is the ultimate goal of the proposed methodology: in this context, only a subset of the space is of interest, \ie where the cost function is minimal. To take advantage of this, we consider a two-stage enrichment scheme as originally proposed in \citet{MoustaphaSMO2016}. In the first stage, enrichment is done as described in Case $\#2$ but with a looser convergence criterion. Herein, the thresholds $\epsilon_{D_f}$ and $\epsilon_{SC}$ in Eqs.~\eqref{eq:enr_conv1} and \eqref{eq:enr_conv2} are both increased to $0.01$. The idea behind is to locate the limit-state surface without necessarily focusing on extreme accuracy. The remaining epistemic uncertainty due to surrogate modelling is taken care of in the second stage where optimization is coupled with enrichment. The procedure consists in ensuring accuracy of the estimated failure probabilities, or herein quantiles, at each iteration of the optimization process. For this, we consider a metric introduced in \citet{Dubourg2011} and adapted for quantile estimation in \citet{MoustaphaSMO2016} where an upper and lower bounds of the quantiles estimates  (resp. $q_{\alpha}^-(\ve{d})$ and $q_{\alpha}^+(\ve{d})$) are  introduced using the models defined respectively by $\mu_{\mathcal{M}^{\text{K}}}\prt{\ve{x}} - 1.96 \, \sigma_{\mathcal{M}^{\text{K}}} \prt{\ve{x}}$ and $\mu_{\mathcal{M}^{\text{K}}}\prt{\ve{x}} + 1.96 \, \sigma_{\mathcal{M}^{\text{K}}} \prt{\ve{x}}$. It can be shown that
\begin{equation}
q_{\alpha}^-(\ve{d}) \leq q_{\alpha}(\ve{d}) \leq q_{\alpha}^+(\ve{d}).
\end{equation}
Even though this bound does not indicate where the true quantile lies, it is a measure of how accurate the estimated quantile is  with respect to the Kriging epistemic uncertainty. Therefore, at each iteration of the optimization process, the following criterion is checked:
\begin{equation}
\frac{\abs{q_{\alpha}^-(\ve{d}) - q_{\alpha}^+(\ve{d})} }{ 1 + \abs{q_{\alpha}(\ve{d})} } \leq \epsilon_q,
\end{equation}
where $\epsilon_q$ is a threshold to be defined. If this condition is not respected, then enrichment is made locally using the procedure introduced in Section~\ref{sec:AK} where the candidates for enrichment are simply the set of points that are locally used to compute the failure probability.

For this special application using CMA-ES, the criterion is computed only for sampled points that are feasible and improve the current best point. This is to avoid enrichment in the exploratory part of CMA-ES. Other tricks can be used to make the procedure more robust and efficient (\eg a threshold $\epsilon_q$ that decreases as the optimization progresses) but this is not in the scope of this paper. With this slight modification, Case $\#3$ does not exactly illustrate the proposed framework and is qualified here as \emph{gray-box} because of the coupling between optimization and surrogate modelling. Its main interest however is that it greatly enhances the efficiency of the solution.

\section{Application examples}\label{sec:Applications}
In this section, the three configurations of the proposed framework are illustrated using three examples of increasing complexity. Furthermore, the versatility of the surrogate framework is shown by considering the different classes of probabilistic inputs which consist in a) design variables with uncertainty $\ve{X}\prt{\ve{d}}$ only, b) design and environmental variables and c) deterministic design parameters and random environmental variables. 
\subsection{Two-dimensional highly non-linear limit-state function}
In this first example, we consider a two-dimensional highly non-linear function for which the RBDO problem reads \citep{Lee2008}:
\begin{equation}\label{eq:031}
\begin{split}
 \ve{d}^\ast = \arg \min_{\ve{d} \in \bra{0, \,3.7} \times \bra{0, \, 4}} & (d_1 - 3.7)^2 + (d_2 - 4)^2  \\
  \text{subject to : } & \left\{ \begin{array}{ll}
\displaystyle{\mathfrak{f}\prt{\ve{d}} =3 - d_1 - d_2  \leq 0,} \\
\displaystyle{\Prob{\mathfrak{g} \prt{\ve{X}\prt{\ve{d}}} =X_1 \, \sin\prt{4 \, X_1} + 1.1 \, X_2 \, \sin(2 \, X_2) \leq 0} \leq \Phi(-2)},
\end{array} \right.
\end{split}
\end{equation}
where $X_i \sim \mathcal{N}\prt{d_i, \, 0.1^2}, i = \acc{1,2}$. The reference solution for this problem is $\mathfrak{c}^\ast = 1.3285$ corresponding to an optimal design $\ve{d}^\ast = \acc{2.8582, 3.2127}$. This solution is found using a direct double-loop approach on the original functions with a large size Monte Carlo simulation in the inner loop (see also \citet{DubourgThesis}).

\begin{figure}
	\centering
	\subfloat[Optimal cost]{\label{fig:Ex1_Results_a}\includegraphics[width=0.49\textwidth]{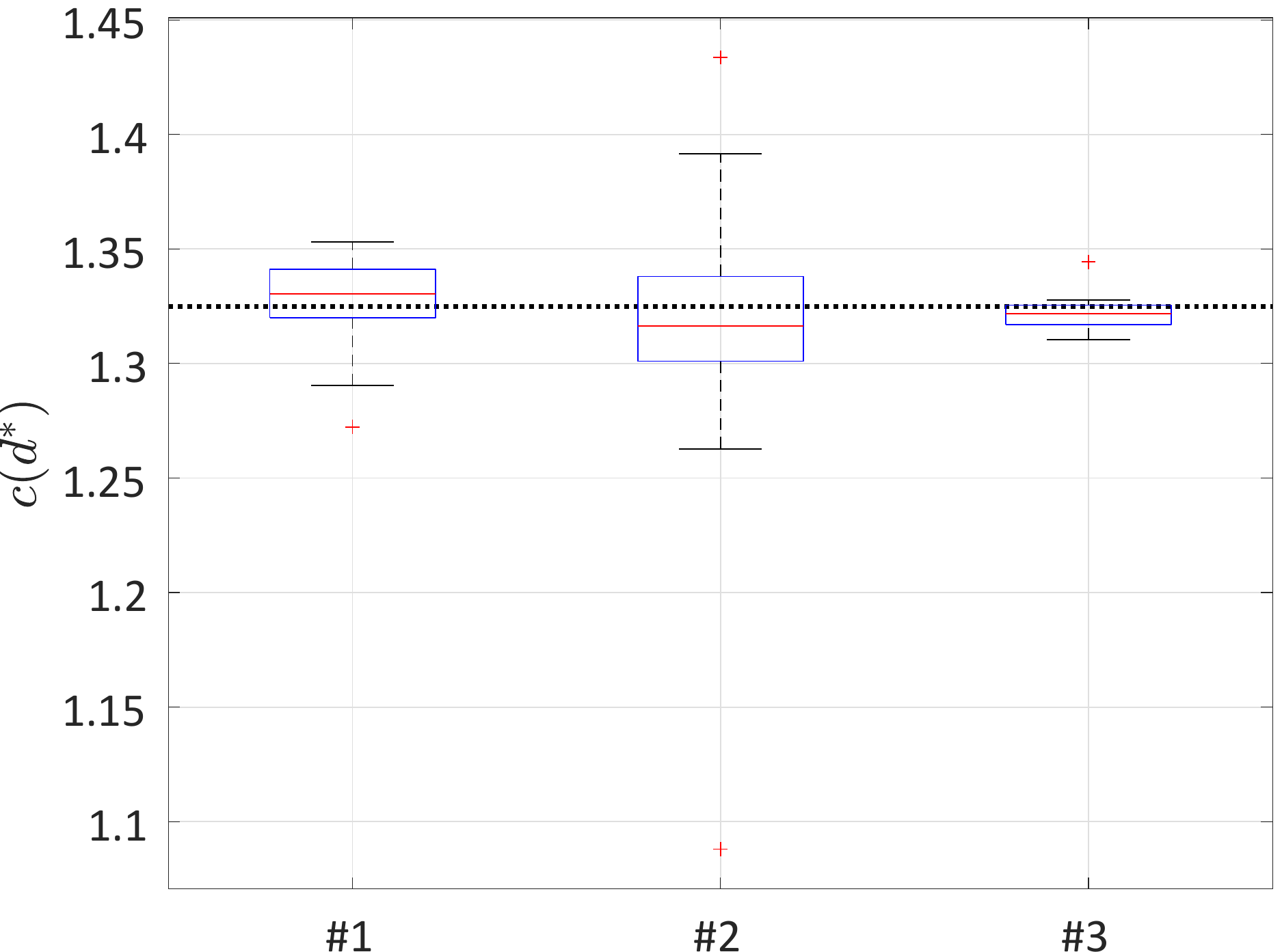}}%
	\hfill
	\subfloat[Model evaluations]{\label{fig:Ex1_Results_b}\includegraphics[width=0.49\textwidth]{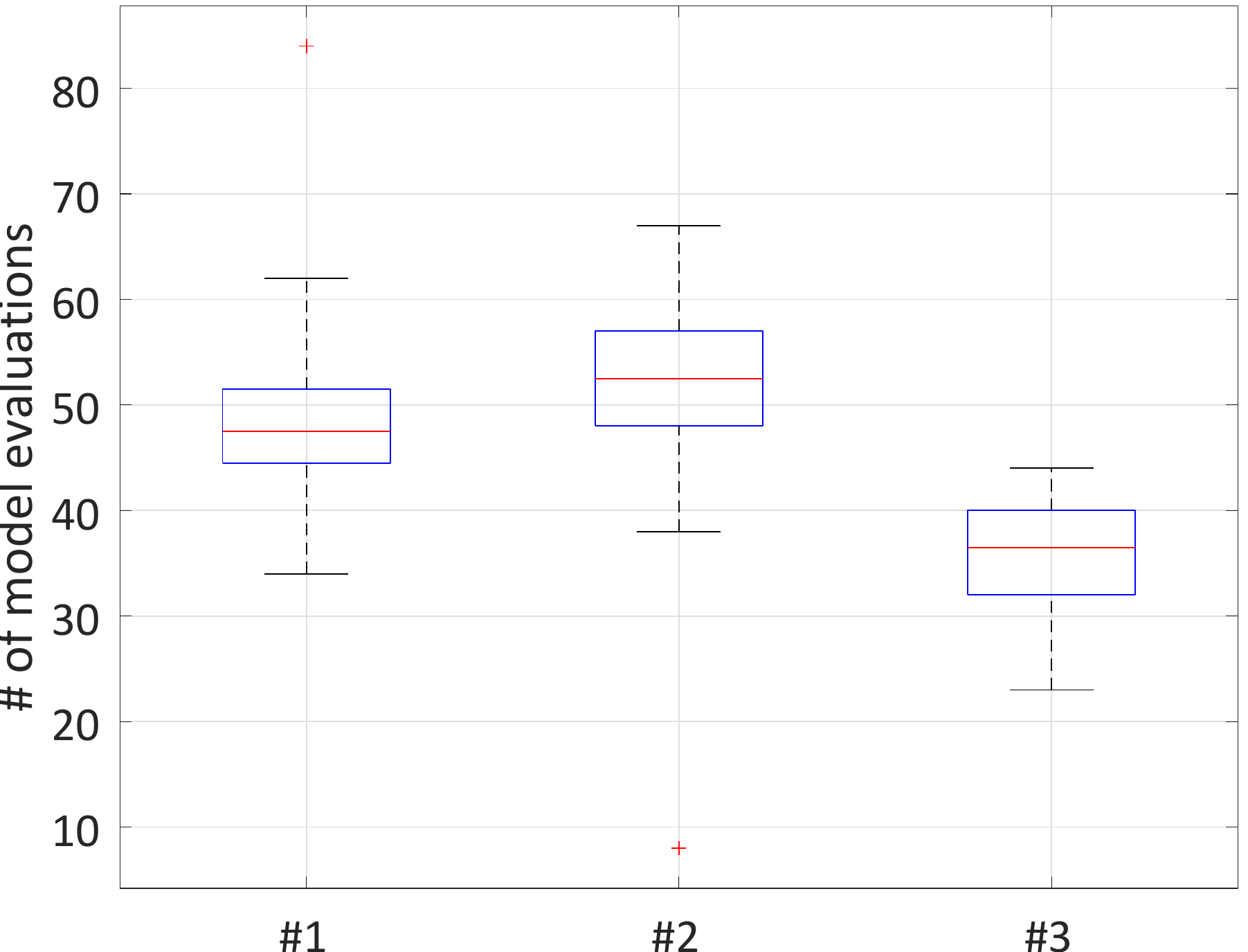}}%
	\caption{Example 1 - Two-dimensional highly non-linear limit-state function: optimized cost and number of model evaluations for the three strategies (box-plots summarizing $20$ replications).}
	\label{fig:Ex1_Results}
\end{figure}

To solve this problem, an initial experimental design of size $6$ is drawn using Latin hypercube sampling (LHS). The solution is repeated $20$ times to account for the inherent statistical variability of the different solution schemes. Figure~\ref{fig:Ex1_Results} shows boxplots of the found solutions (left panel) together with the total number of model evaluations (right panel) for each case. The results are also summarized in \tabref{tab:Ex1_results} together with some benchmark solutions for comparison purposes. It can be observed that the three framework configurations lead to an accurate solution at a reasonable cost. In fact the number of calls to the original model is smaller than approximation approaches such as PMA as reported by \citet{Lee2008} or the Kriging-based RBDO approach proposed in \citet{Dubourg2011}. The larger number reported in \citet{Dubourg2011} can be explained by the fact that they also approximate the first constraint by a surrogate model.
\begin{table}[!ht]
	\centering
		\begin{threeparttable}
	\caption{Example 1 - Two-dimensional highly non-linear limit-state function: optimization results and comparison of costs.}
	\label{tab:Ex1_results}
	\begin{tabular}{lcccc}
		\hline
		{Method}    & {$d_1^\ast$}  & {$d_2^\ast$}  & {$\mathfrak{c}\prt{\ve{d}}^\ast$} & {$N$}\\ \hline
		{Reference solution}		  & 	{$2.85$}		& 	{$3.23$}	   & 	{$1.33$}	&	{$\sim 10^7$}	   \\
		{Performance measure approach (PMA)$^a$}		      & 	{$2.84$}		& 	{$3.23$}	   & 	{$1.33$}	&	{$296$}	   \\
		{Meta-RBDO$^b$}	  	  & 	{$2.81$}	  	& 	{$3.25$}  	   & 	{$1.35$}	&	{$80$}   \\
		{Framework $\#1$$^c$} & 	{$2.87$}	  	& 	{$3.20$}  	   & 	{$1.33$}	&	{$47.5$}   \\
		{Framework $\#2$$^c$} & 	{$2.88$}	  	& 	{$3.20$}  	   & 	{$1.32$}	&	{$52.5$}   \\
		{Framework $\#3$$^c$} & 	{$2.84$}	  	& 	{$3.24$}  	   & 	{$1.32$}	&	{$36.5$}   \\
		\hline
	\end{tabular}
		\begin{tablenotes}
	\tiny
	\item $^a$ As calculated in \citet{Lee2008}
	\item $^b$ As calculated in \citet{DubourgThesis}
	\item $^c$ Median values found from $20$ replications
\end{tablenotes}
	\end{threeparttable}
\end{table}

\begin{figure}[!ht]
	\centering
	\subfloat[Refererence]{\label{fig:Ex1_Enr_a}\includegraphics[width=0.49\textwidth]{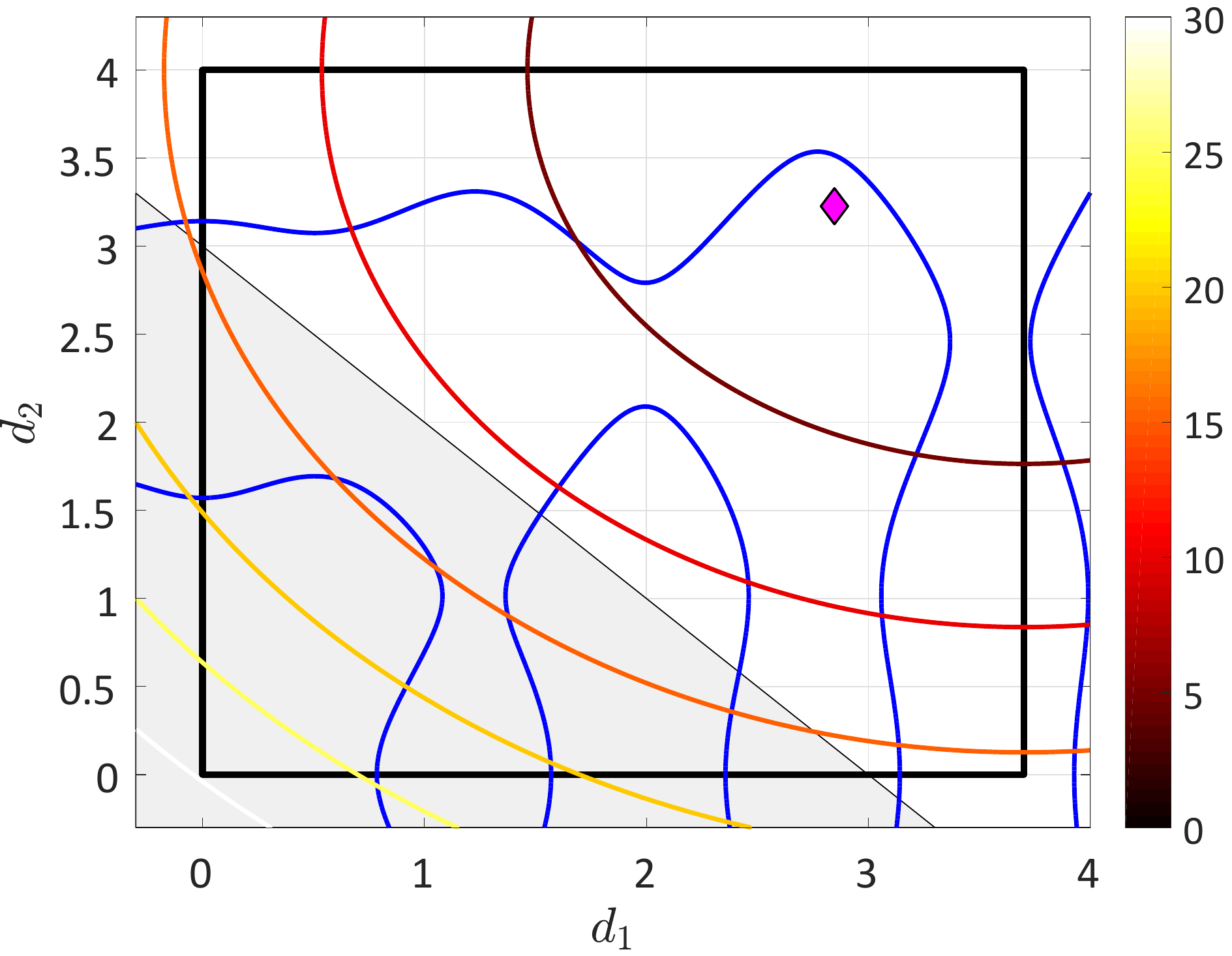}}%
	\hfill
	\subfloat[Case $\#1$]{\label{fig:Ex1_Enr_b}\includegraphics[width=0.49\textwidth]{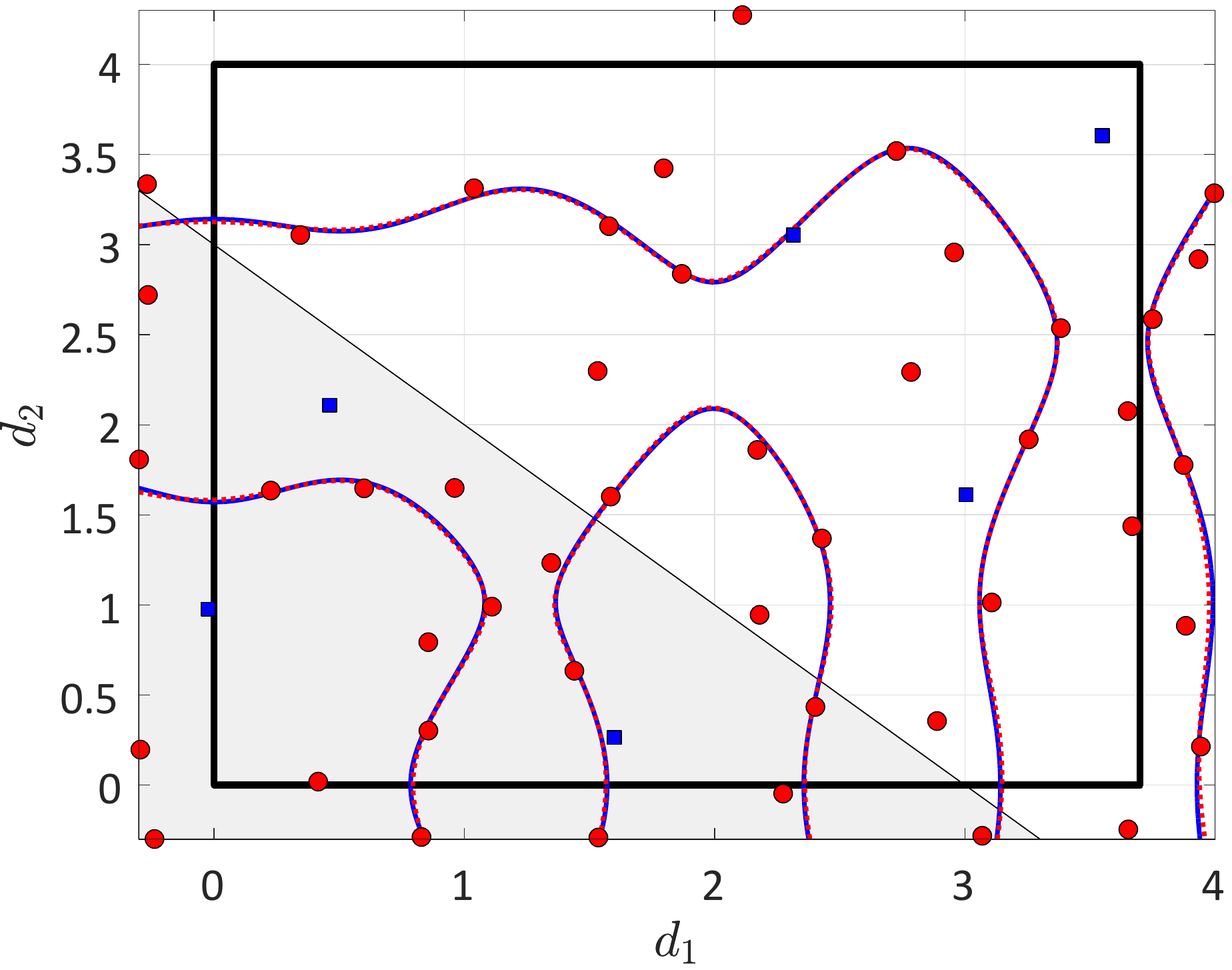}}%
	\\
	\subfloat[Case $\#2$]{\label{fig:Ex1_Enr_c}\includegraphics[width=0.49\textwidth]{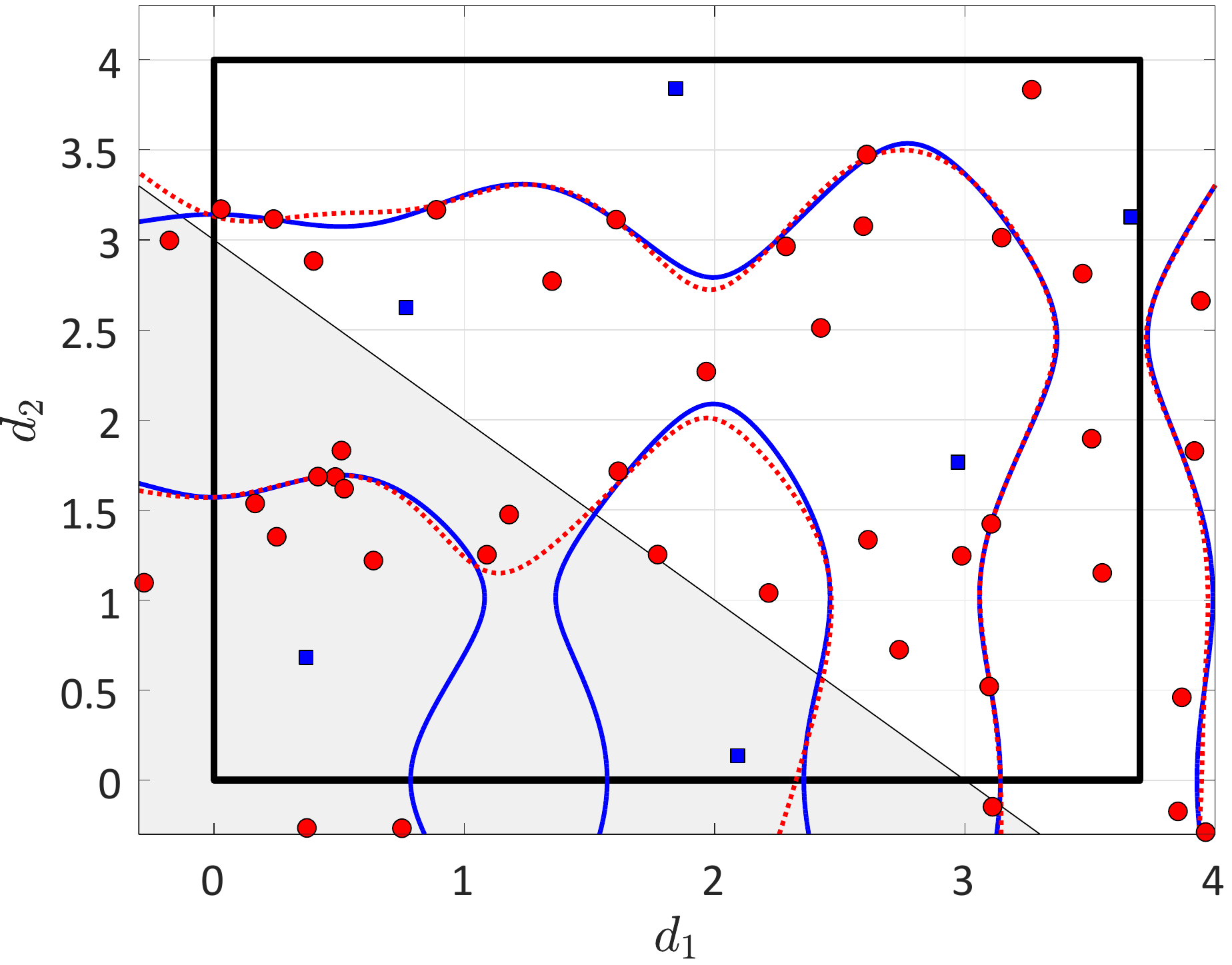}}%
	\hfill
	\subfloat[Case $\#3$]{\label{fig:Ex1_Enr_d}\includegraphics[width=0.49\textwidth]{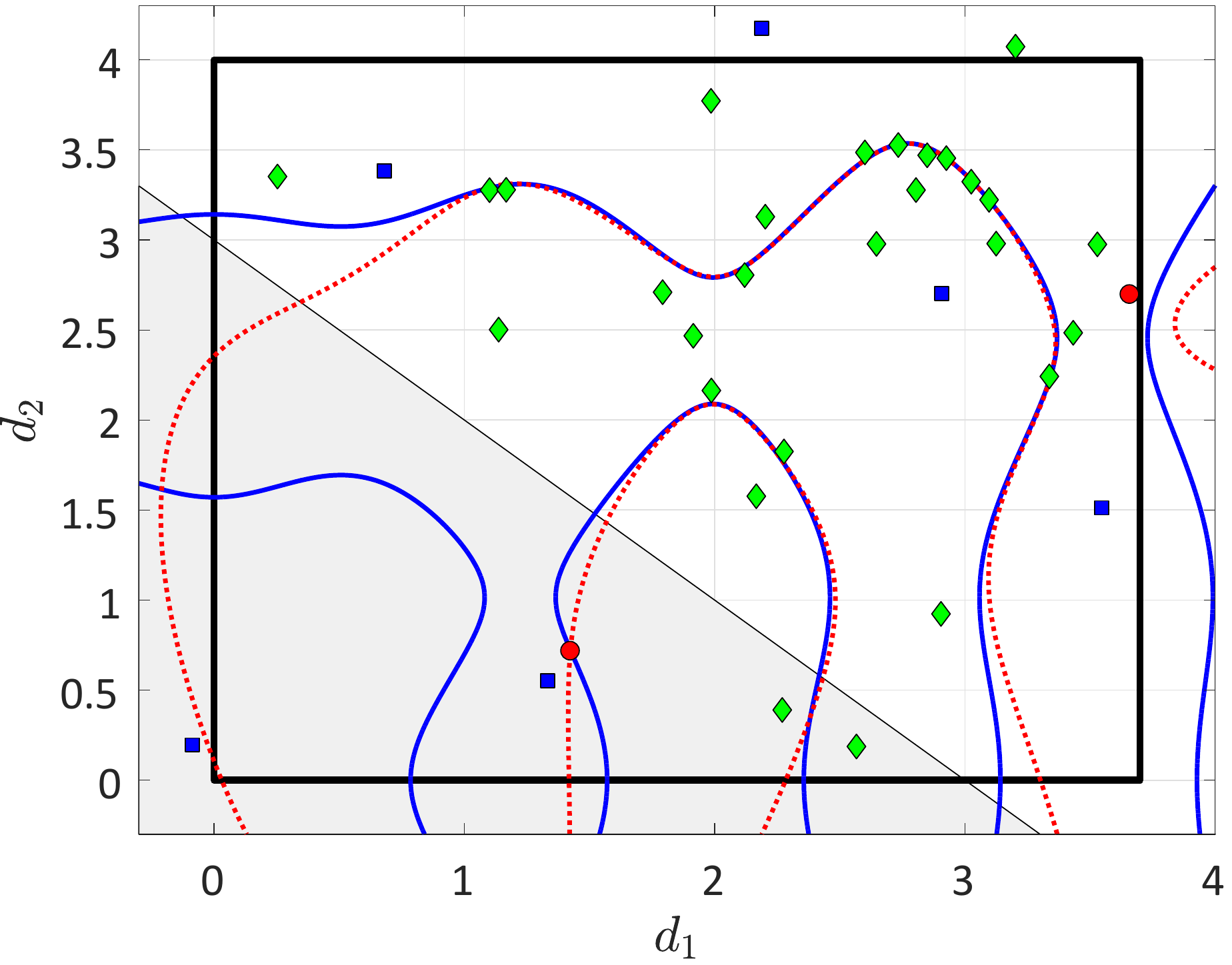}}%
	\caption{Example 1 - Two-dimensional highly non-linear limit-state function: initial experimental design and enrichment points for the the median solution (out of $20$ replications) for the three proposed solving strategies.}
	\label{fig:Ex1_Enr}
\end{figure}

Let us now look at the convergence of each framework configuration. Figure~\ref{fig:Ex1_Enr} shows the final surrogate models and the experimental design points for the median solution of each solving strategy. The entire panel window shows the augmented space while the design space is limited by the inner black rectangle. As explained earlier, the augmented space in this case is simply obtained by extending the bounds of the design space in each direction. The limit-state surface as computed by the original model is drawn with the thick blue line and the gray shaded area represents the unfeasible space due to the soft constraint. Figure~\ref{fig:Ex1_Enr_a} show additionally contours of the cost function and the reference solution represented by the magenta diamond.  In the remaining panels, the initial experimental design is shown by the blue squares. The red circles correspond to the enrichment points added prior to starting the optimization process. The approximated limit-state surface is represented by the dashed red line. The cases $\#1$ and $\#2$ show approximately the same enrichment scheme. The case where SVR together with the maximin enrichment scheme are used (Figure~\ref{fig:Ex1_Enr_b}) produces the most uniform sampling with almost all the new points located near to the final limit-state surface. The framework configuration $\#2$ (Figure~\ref{fig:Ex1_Enr_c}) produces a slightly less accurate Kriging approximation. The case $\#3$ (Figure~\ref{fig:Ex1_Enr_d}) corresponds to the coupling of enrichment with optimization. In fact, only two points are added at the first stage of enrichment. The green diamonds show the points that are added during optimization. Almost all the points added in this stage are located around the solution. The final Kriging model is not accurate everywhere in the space but only in areas of the limit-state surface around the optimum. Overall, this allows the approach to converge to the solution with fewer number of model evaluations. Particularly, very few points are added where the cost function is large or in unfeasible regions due to the soft constraints.

\subsection{Short column under oblique bending} 
This  benchmark deals with a short column structure of rectangular cross-section $b \times h$ which is subject to an axial load $F$ and biaxial bending moments $M_1$ and $M_2$. The limit-state function characterizes the performance of the structure with respect to its yield stress $\sigma_y$ and reads:
\begin{equation}
\mathfrak{g}\prt{\ve{X}\prt{\ve{d}},\ve{Z}}  = 1 - \frac{4 M_1}{b h^2 \sigma_y} - \frac{4 M_2}{b^2 h \sigma_y} - \prt{\frac{F}{b h \sigma_y}}^2.
\end{equation}
The probabilistic model is defined in Table~\ref{tab:column_param}. Note that the uncertainty in the design parameters has been artificially introduced by \citet{DubourgThesis} to accommodate his method. Even though in the proposed framework such trick is not necessary, we keep the same problem formulation so as to be able to compare the results.
\begin{table}[!ht]
	\centering
	\caption{Example 2 - Short column under oblique bending: parameters of the variables defining the probabilistic model - $\ve{d} = \acc{b, h}^T$ are the design variables and $\ve{Z} = \acc{F, M_1, M_2, \sigma_y}^T$ are the environmental variables.}
	\label{tab:column_param}
	\begin{tabular}{lccc}
		\hline
		{Parameter}    & {Distribution}  & {Mean}  & {COV ($ \delta \%$)} \\ \hline
		{$b$ (mm)}		      & {Normal}		& {$\mu_b$}	   & {$0.01$}	   \\
		{$h$ (mm)}	  & {Normal}	  & {$\mu_h$}  & {$0.01$}   \\
		{$F$ (N)}		  &{Lognormal}	  & {$2.5 \times 10^6$}	 & {$0.20$}    \\
		{$M_1$ (N.mm)}		  &{Lognormal}	  & {$250 \times 10^6$}	 & {$0.30$}    \\
		{$M_2$ (N.mm)}		  &{Lognormal}	  & {$125 \times 10^6$}	 & {$0.30$}    \\
		{$\sigma_y$ (MPa)}		  &{Lognormal}	  & {$2.5 \times 10^6$}	 & {$0.10$}    \\
		\hline
	\end{tabular}
\end{table}

The optimization is started with an initial experimental design of size $21$. For this six-dimensional example, the convergence criteria do not lead to robust solutions with respect to the replications. As shown in Table~\ref{tab:Ex2_results}, the median number of calls to the limit-state surface are respectively $84$ and $86$ for the cases $\#1$ and $\#2$. The corresponding solutions however show some scatter around the reference value. The case $\#3$ allows us to reduce both the solution scatter and the average number of model evaluations. 
\begin{figure}
	\centering
	\subfloat[Optimal cost]{\label{fig:Ex2_Results_a}\includegraphics[width=0.49\textwidth]{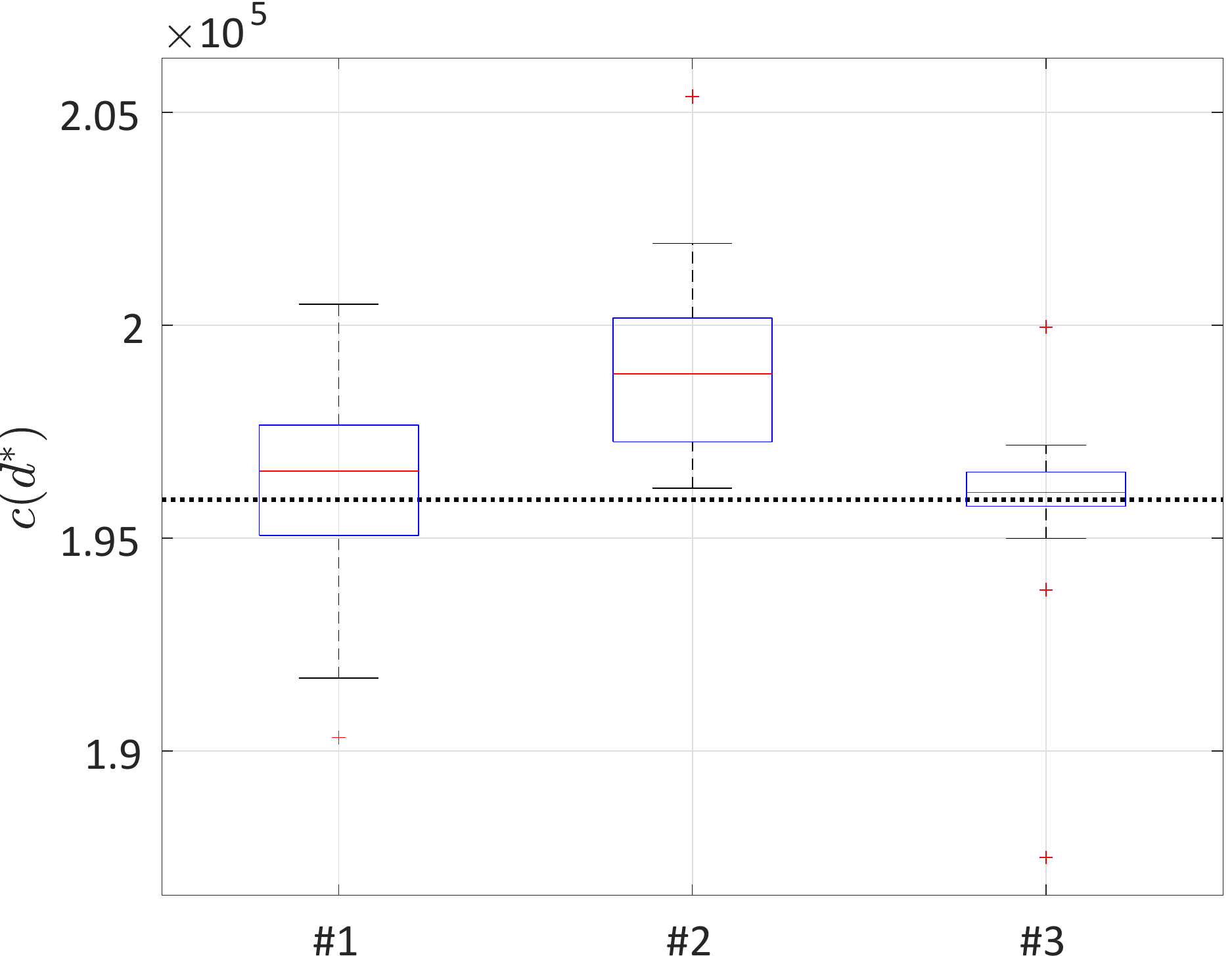}}%
	\hfill
	\subfloat[Model evaluations]{\label{fig:Ex2_Results_b}\includegraphics[width=0.49\textwidth]{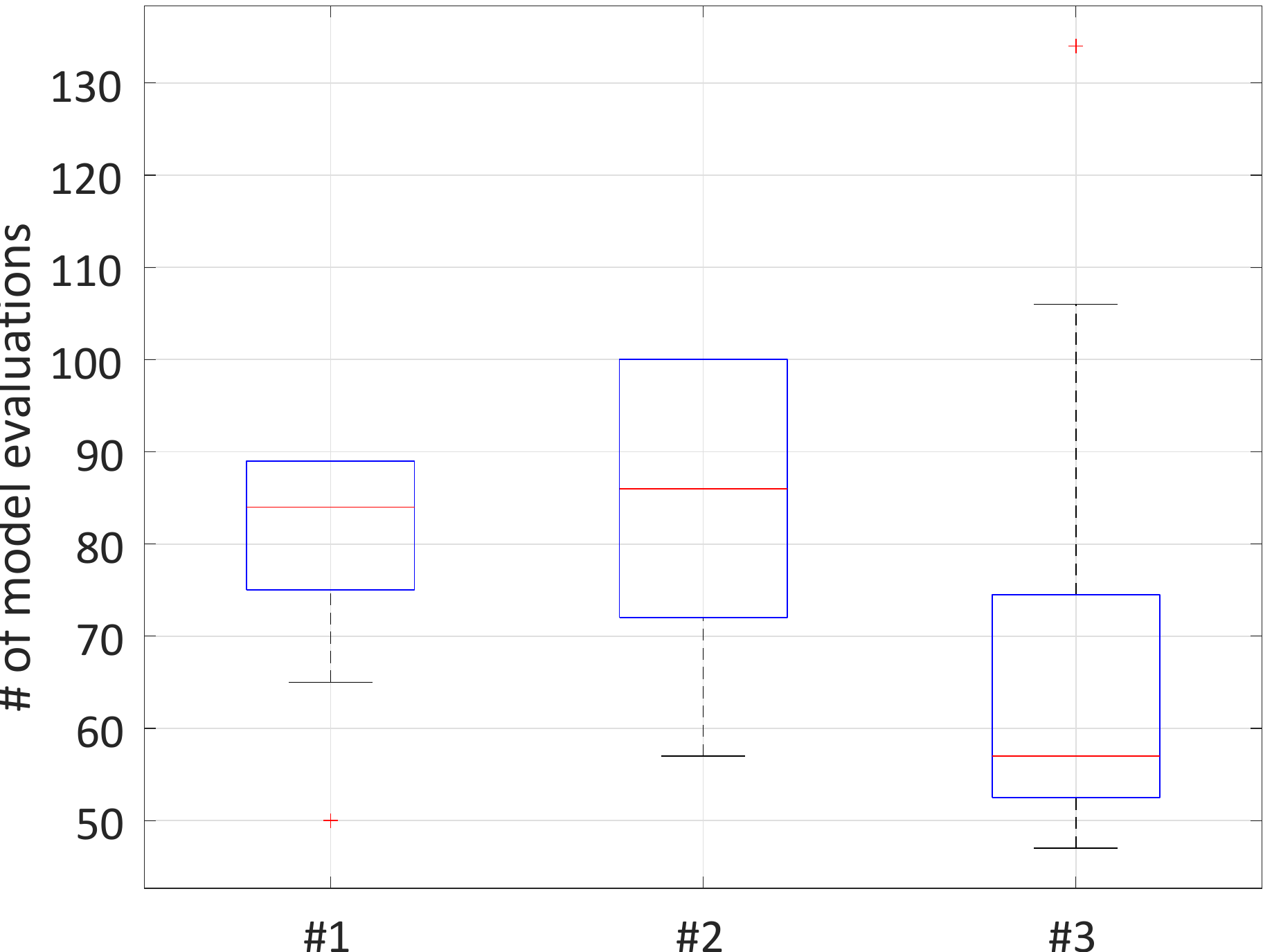}}%
	\caption{Example 2 - Short column under oblique bending: optimized cost and number of model evaluations for the three strategies (box-plots summarizing $20$ replications).}
	\label{fig:Ex2_Results}
\end{figure}
\begin{table}[!ht]
	\centering
	\begin{threeparttable}
		\caption{Example 2 - Short column under oblique bending: optimization results and comparison of costs.}
		\label{tab:Ex2_results}
		\begin{tabular}{lccccc}
			\hline
			{Method}    & {$d_1^\ast$}  & {$d_2^\ast$}  & {$\mathfrak{c}\prt{\ve{d}}^\ast$} & {$N$} & {$\beta^\dagger$} \\\hline
			{Reference solution}		  & 	{$334$}		& 	{$587$}	   & 	{$1.96 \times 10^5 $}	&	{$\sim 10^7$}	  & {$3.00$} \\
			{Nested FORM$^a$}		      & 	{$399$}		& 	{$513$}	   & 	{$2.12 \times 10^5 $}	&	{$9,472$}	& {$3.38$}   \\
			{Meta-RBDO$^b$}	  	  & 	{$358$}	  	& 	{$580$}  	   & 	{$2.15 \times 10^5 $}	&	{$70$}   & {$3.32$} \\
			{Framework $\#1$$^c$} & 	{$332$}	  	& 	{$596$}  	   & 	{$1.97 \times 10^5 $}	&	{$84$}   & {$3.00$} \\
			{Framework $\#2$$^c$} & 	{$336$}	  	& 	{$592$}  	   & 	{$1.99 \times 10^5 $}	&	{$86$}   & {$3.00$} \\
			{Framework $\#3$$^c$} & 	{$331$}	  	& 	{$592$}  	   & 	{$1.96 \times 10^5 $}	&	{$57$}   & {$3.00$} \\
			\hline
		\end{tabular}
		\begin{tablenotes}
			\tiny
			\item $^a$ As calculated in \citet{Lee2008}
			\item $^b$ As calculated in \citet{DubourgThesis}
			\item $^c$ Median values found from the $20$ replications
			\item $^\dagger$ The constraints are not saturated in the other references because their RBDO formulation is slightly different: $\ve{d} = \arg \min_{\ve{d}} \mathfrak{c}\prt{\ve{d}} + C_f P_f\prt{\ve{X}\prt{\ve{d}}, \ve{Z}}$
		\end{tablenotes}
	\end{threeparttable}
\end{table}

\subsection{Dome structure}
\begin{figure}[!ht]
	\begin{center}
		\includegraphics[width=0.66\textwidth]{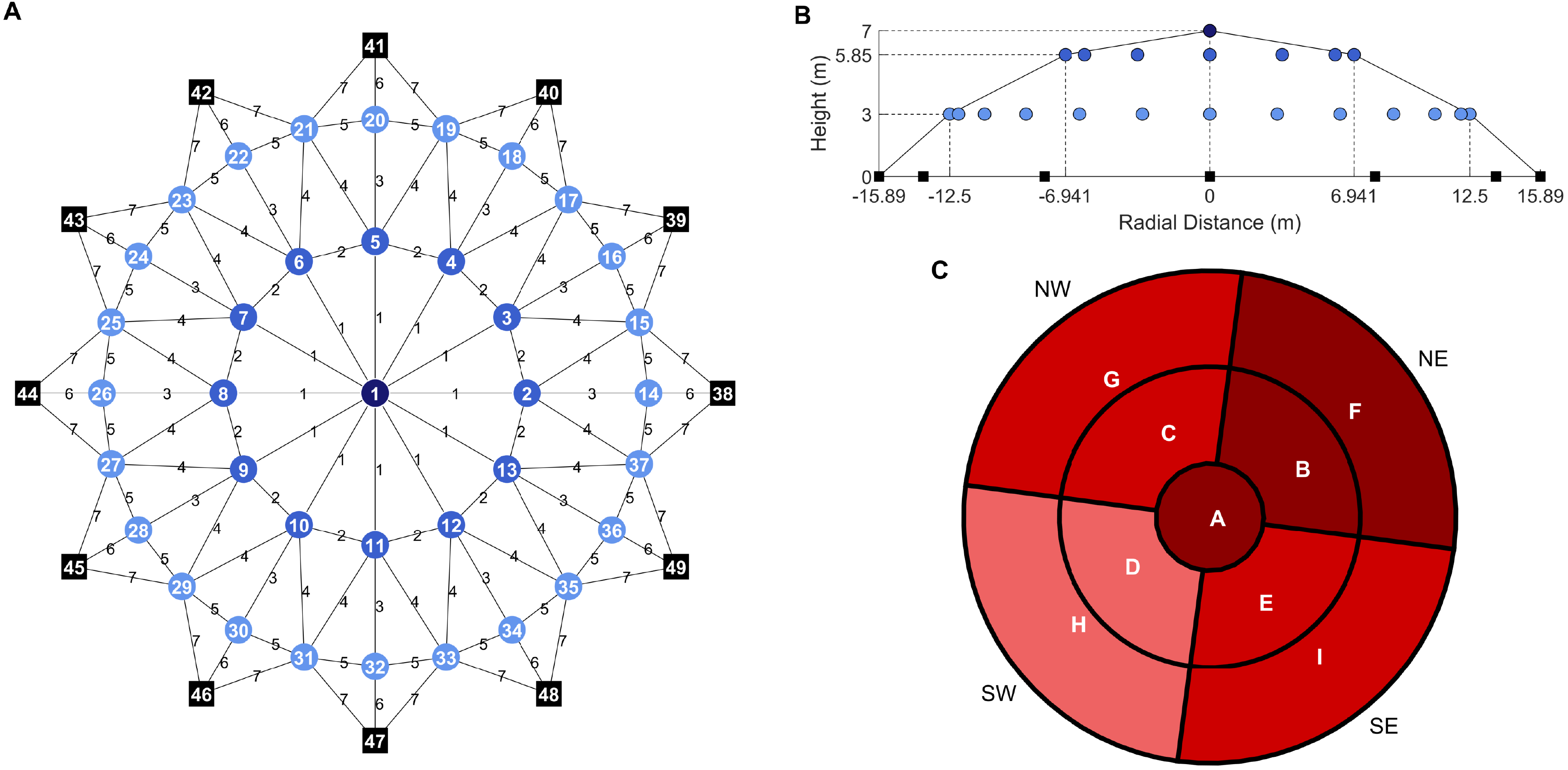}
	\end{center}
	\caption{Example 3 - Dome structure: geometry and loading sectors (adapted from \citet{Torre2018_Vines4UQ})}
	\label{fig:dome}
\end{figure}

This final example is related to the optimization of the dome structure introduced by \citet{Kaveh2009}, albeit in a deterministic context. This problem is extended here to account for uncertainties in the loading as proposed by \citet{Torre2018_Vines4UQ}. The dome, illustrated in Figure~\ref{fig:dome}, consists of $120$ bars divided into seven groups as numbered in panel A. The dome surface is divided into seven sectors to which a Gumbel-distributed load is assigned with the following mean values: $1$ kN/m$^2$ in the top and north-east sectors A, B and F, $0.5$ kN/m$^2$ in the north-west and south-east sectors C, E, G and I and finally $0.25$ kN/m$^2$ in the south-west sectors D and H. The coefficient of variation is $0.2$ in all cases. These loads are supposed to model snow falling from north-east direction \citep{Torre2018_Vines4UQ}. In addition, self-weight and deterministic service loads are applied to each node of the structure as follows: $60$~kN to node $1$, $30$~kN on nodes $2$ to $13$ and $10$~kN on nodes $14$ to $37$ \citep{Kaveh2009}. The node groups and sectors are illustrated using different color schemes (respectively shades of blue and red).

The optimization problem consists in minimizing the weight of the structure with design variables being the bars cross-sectional areas $\ve{d} \in \mathbb{D} = \bra{10, \, 40}$ cm$^2$ while ensuring that the vertical displacement of node $\#2$ remains below a given threshold $\bar{\delta} = 10$ cm. Accounting for uncertainties, the target failure probability is set to $\bar{P}_f = 0.01$.

\begin{figure}[!ht]
	\begin{center}
		\includegraphics[width=0.4\textwidth]{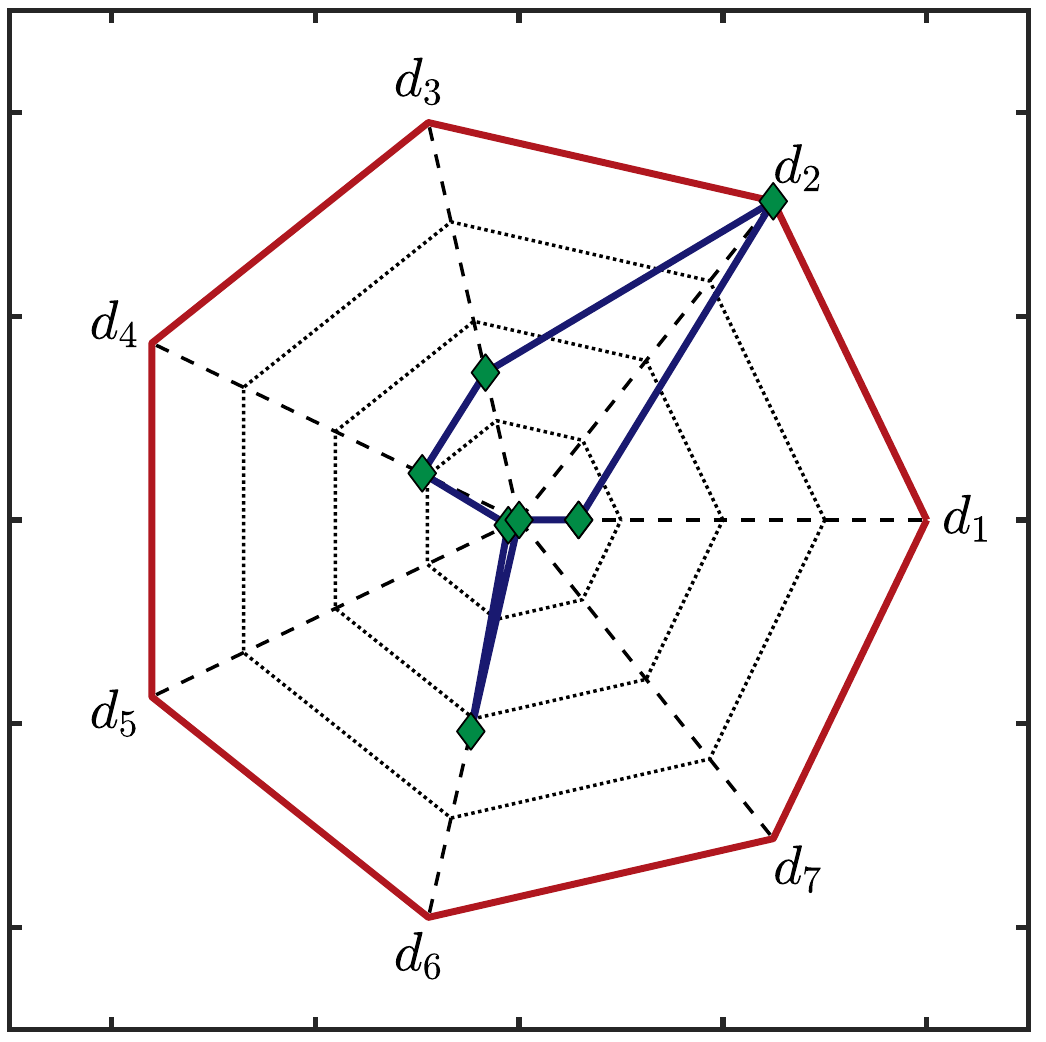}
	\end{center}
	\caption{Relative coordinates of the optimal design with respect to their respective lower and upper bounds.}
	\label{fig:Dome_Xstar}
\end{figure}

The problem is solved using only the framework configuration $\#3$ which has proven to be the most efficient in the other applications. An initial experimental design of $80$ points is drawn using optimal Latin hypercube sampling. The first stage of enrichment is then carried out and convergence is achieved after two iterations only. The final solution is achieved with a total of $154$ model evaluations. The optimal weight is found to be $8.68$ tons. Figure~\ref{fig:Dome_Xstar} shows the relative areas with respect to lower and upper bounds for each design dimension. The largest section is attributed to the group of bars labeled $2$ which is close to the maximum allowable size. The two groups of bars which are not in the radial direction and not connected to node $\#2$ do not really influence the deflection and are hence set close to their lower bounds by the optimizer. 

\begin{figure}[!ht]
	\centering
	\includegraphics[width=0.49\textwidth]{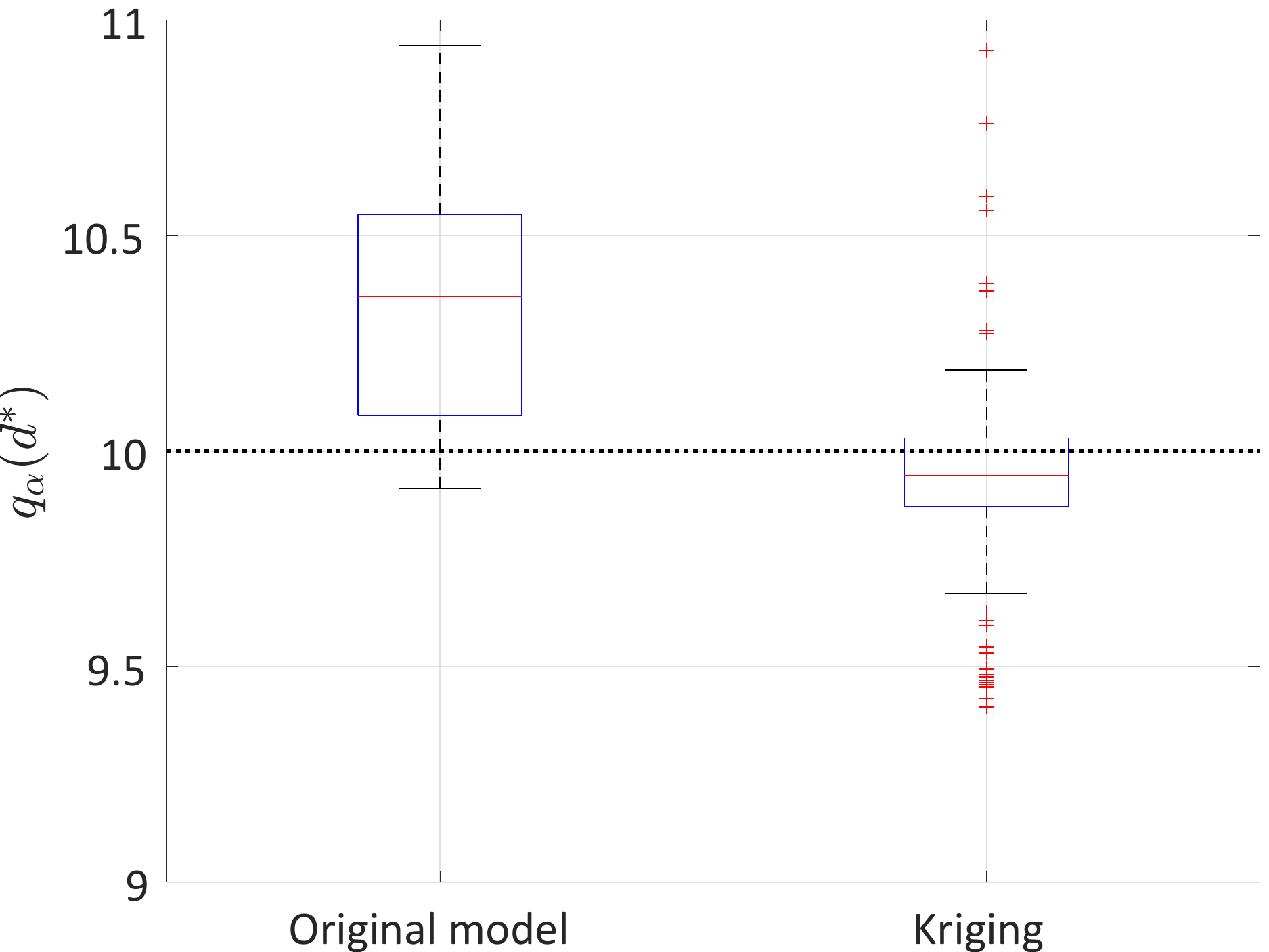}%
	\caption{Example 3 - Dome structure: Comparison of the quantiles at th efinal design computed using the original model and the final Kriging model.}
	\label{fig:Ex3_boxplot}
\end{figure}

This result given above assumes that the surrogate model is accurate enough. In order to validate the surrogate model accuracy in the vicinity of the limit-state surface, a Monte Carlo set of $1,000$ points is drawn using the final design, \ie the design components are those of the optimal design while the environmental variables follow their own distributions. Since the constraints are saturated, the estimated quantiles should be equal to $\bar{\delta} = 10$ cm. Figure~\ref{fig:Ex3_boxplot} shows the boxplot of the quantiles with $500$ bootstrap replications when using the final Kriging model and the original model. Bootstrap is used here to account for the small size of the validation Monte Carlo set. Accounting for the bootstrap standard deviation, the targeted threshold belongs to a small confidence interval around the estimated quantiles. Thus the final Kriging model is deemed accurate enough considering this validation set even though it seems to slightly underestimate the true quantile. Note that a better fit could be obtained by calibrating the enrichment convergence criterion to allow for more enrichment points.

Finally it is worth emphasizing that only $154$ finite element runs were used to solve a $16$-dimensional RBDO problem.

\section{Conclusion}
This paper aims at introducing a global and unified framework for the solution of reliability-based design optimization (RBDO) problems. A review of classical approximation-based approaches is first carried out following the widely accepted classification into double-loop, single-loop and decoupled methods similar to \citet{Chateauneuf2008b}. Upon stressing the limitations of such methods, more recent approaches that combine simulation techniques for the reliability analysis and surrogate-modelling are further reviewed. These methods have been developed mainly since 2010. They are classified here according to the way the surrogate model is integrated into the RBDO framework. 

Then, a global framework that combines three independent blocks, namely adaptive surrogate modelling, reliability analysis and optimization, is proposed together with some numerical considerations, \eg{} on the gradients estimation. It is argued that the proposed framework is modular and non-intrusive, meaning that each block can be set independently from the others. Hence the analyst can freely choose his favorite surrogate modelling, reliability analysis and optimization method.

Finally, an illustration of the proposed framework is made using different combinations for each block. Each of the three examples treats an RBDO formulation with the three classes of probabilistic inputs: only random design parameters which appears as parameters of random variables; deterministic design- and random environmental variables and eventually the combination of all types of variables. All these cases can be treated in the same way through the use of an augmented space. It is worth emphasizing that most of the current literature focus on one formulation and require adaptations to treat other scenarios. The results for the three examples are compared to literature references. Efficiency of the framework is demonstrated by the number of calls to the performance functions, which is in the same range as other surrogate-based methods in the literature. As expected, the efficiency and robustness is much larger than approaches that rely on approximation methods such as FORM for the reliability analysis. However results are dependent on the convergence of the enrichment scheme which require a proper calibration. The proposed general framework is currently integrated as a new module \citep{UQdoc_12_114} of the uncertainty quantification platform \textsc{UQLab} \citep{MarelliUQLab2014}, which will ensure further dissemination of surrogate-based RBDO methods.
%%%%%%%%%%%%%%%%%%%%%%%%%%%%%%%%%%%%%%%%%%%%%%%%%%%%%%%%%%%%%%%%%%%%%%%%%%%

%% References
\bibliographystyle{chicago}
\bibliography{biblioRSUQ}

\begin{thebibliography}{}

\bibitem[\protect\citeauthoryear{Agarwal, Mozumder, Renaud, and Watson}{Agarwal
  et~al.}{2007}]{Agarwal2007}
Agarwal, H., C.~K. Mozumder, J.~E. Renaud, and L.~T. Watson (2007).
\newblock An inverse-measure-based unilevel architecture for reliability-based
  design optimization.
\newblock {\em Struct. Multidisc. Optim.\/}~{\em 33\/}(3), 217--227.

\bibitem[\protect\citeauthoryear{Agarwal and Renaud}{Agarwal and
  Renaud}{2004}]{Agarwal2004}
Agarwal, H. and J.~Renaud (2004).
\newblock Reliability-based design optimization using response surfaces in
  application to multidisciplinary systems.
\newblock {\em Eng. Opt.\/}~{\em 36\/}(3), 291--311.

\bibitem[\protect\citeauthoryear{Aoues and Chateauneuf}{Aoues and
  Chateauneuf}{2010}]{Aoues2010}
Aoues, Y. and A.~Chateauneuf (2010).
\newblock Benchmark study of numerical methods for reliability-based design
  optimization.
\newblock {\em Struct. Multidisc. Optim.\/}~{\em 41\/}(2), 277--294.

\bibitem[\protect\citeauthoryear{Arnold and Hansen}{Arnold and
  Hansen}{2012}]{Arnold2012}
Arnold, D.~V. and N.~Hansen (2012).
\newblock {A (1+1)-CMA-ES for constrained optimisation}.
\newblock In T.~Soule and J.~H. Moore (Eds.), {\em Genetic and evolutionary
  computation conference}, pp.\  297--304.

\bibitem[\protect\citeauthoryear{Au}{Au}{2005}]{Au2005}
Au, S.-K. (2005).
\newblock Reliability-based design sensitivity by efficient simulation.
\newblock {\em Comput. Struct.\/}~{\em 83\/}(14), 1048--1061.

\bibitem[\protect\citeauthoryear{Au and Beck}{Au and Beck}{2001}]{Au2001}
Au, S.~K. and J.~L. Beck (2001).
\newblock Estimation of small failure probabilities in high dimensions by
  subset simulation.
\newblock {\em Prob. Eng. Mech.\/}~{\em 16\/}(4), 263--277.

\bibitem[\protect\citeauthoryear{Bachoc}{Bachoc}{2013}]{Bachoc2013b}
Bachoc, F. (2013).
\newblock {Cross validation and maximum likelihood estimations of
  hyper-parameters of Gaussian processes with model misspecifications}.
\newblock {\em Comput. Stat. Data Anal.\/}~{\em 66}, 55--69.

\bibitem[\protect\citeauthoryear{Basudhar and Missoum}{Basudhar and
  Missoum}{2008}]{Basudhar2010}
Basudhar, A. and S.~Missoum (2008).
\newblock An improved adaptive sampling scheme for the construction of explicit
  boundaries.
\newblock {\em Struct. Multidisc. Optim.\/}~{\em 42\/}(4), 517--529.

\bibitem[\protect\citeauthoryear{Beaurepaire, Jensen, Schu\"eller, and
  Valdebenito}{Beaurepaire et~al.}{2013}]{Beaurepaire2013}
Beaurepaire, P., H.~A. Jensen, G.~I. Schu\"eller, and M.~A. Valdebenito (2013).
\newblock Reliability-based optimization using bridge importance sampling.
\newblock {\em Prob. Eng. Mech.\/}~{\em 34}, 48--57.

\bibitem[\protect\citeauthoryear{Beck and Gomes}{Beck and
  Gomes}{2012}]{Beck2012}
Beck, A.~T. and W.~J.~S. Gomes (2012).
\newblock A comparison of deterministic, reliability-based and risk-based
  structural optimization under uncertainty.
\newblock {\em Prob. Eng. Mech.\/}~{\em 28}, 18--29.

\bibitem[\protect\citeauthoryear{Bichon, Eldred, Swiler, Mahadevan, and
  McFarland}{Bichon et~al.}{2008}]{Bichon2008}
Bichon, B.~J., M.~S. Eldred, L.~Swiler, S.~Mahadevan, and J.~McFarland (2008).
\newblock Efficient global reliability analysis for nonlinear implicit
  performance functions.
\newblock {\em AIAA Journal\/}~{\em 46\/}(10), 2459--2468.

\bibitem[\protect\citeauthoryear{Bourinet}{Bourinet}{2018}]{BourinetHDR}
Bourinet, J.-M. (2018).
\newblock {\em Reliability analysis and optimal design under uncertainty -
  Focus on adaptive surrogate-based approaches}.
\newblock Universit\'e Blaise Pascal, Clermont-Ferrand, France.
\newblock Habilitation \`a diriger des recherches, 243 pages.

\bibitem[\protect\citeauthoryear{Chapelle, Vapnik, and Bengio}{Chapelle
  et~al.}{2002}]{Chapelle2002}
Chapelle, O., V.~Vapnik, and Y.~Bengio (2002).
\newblock Model selection for small sample regression.
\newblock {\em Machine Learning\/}~{\em 48\/}(1), 9--23.

\bibitem[\protect\citeauthoryear{Chateauneuf}{Chateauneuf}{2008}]{Chateauneuf2008b}
Chateauneuf, A. (2008).
\newblock {\em Structural design optimization considering uncertainties},
  Chapter~1, pp.\  3--30.
\newblock Taylor \& Francis.

\bibitem[\protect\citeauthoryear{Chen, Hasselman, and Neil}{Chen
  et~al.}{1997}]{Chen1997}
Chen, X., K.~Hasselman, T., and D.~J. Neil (1997).
\newblock Reliability-based structural design optimization for practical
  applications.
\newblock In {\em 38th Structures, Structural Dynamics, and Materials
  Conference}, pp.\  2724--2732.

\bibitem[\protect\citeauthoryear{Chen, Peng, Li, Qiu, Xiong, Gao, and Li}{Chen
  et~al.}{2015}]{Chen2015b}
Chen, Z., S.~Peng, X.~Li, H.~Qiu, H.~Xiong, L.~Gao, and P.~Li (2015).
\newblock An important boundary sampling method for reliability-based design
  optimization using {K}riging model.
\newblock {\em Struct. Multidisc. Optim.\/}~{\em 52\/}(1), 55--70.

\bibitem[\protect\citeauthoryear{Cheng, Xu, and Jiang}{Cheng
  et~al.}{2006}]{Cheng2006}
Cheng, G., L.~Xu, and L.~Jiang (2006).
\newblock A sequential approximate programming strategy for reliability-based
  structural optimization.
\newblock {\em Comput. Struct.\/}~{\em 84\/}(21), 1353--1367.

\bibitem[\protect\citeauthoryear{Cho and Lee}{Cho and Lee}{2011}]{Cho2011}
Cho, T.~M. and B.~C. Lee (2011).
\newblock Reliability-based design optimization using convex linearization and
  sequential optimization and reliability assessment method.
\newblock {\em Structural Safety\/}~{\em 33\/}(1), 42--50.

\bibitem[\protect\citeauthoryear{{de Angelis}, Patelli, and Beer}{{de Angelis}
  et~al.}{2015}]{DeAngelis2015}
{de Angelis}, M., E.~Patelli, and M.~Beer (2015).
\newblock Advanced line sampling for efficient robust reliability analysis.
\newblock {\em Structural Safety\/}~{\em 52\/}(B), 170--182.

\bibitem[\protect\citeauthoryear{Ditlevsen and Madsen}{Ditlevsen and
  Madsen}{1996}]{Ditlevsen1996}
Ditlevsen, O. and H.~Madsen (1996).
\newblock {\em Structural reliability methods}.
\newblock {J.~Wiley and Sons, Chichester}.

\bibitem[\protect\citeauthoryear{Du and Chen}{Du and Chen}{2004}]{Du2004}
Du, X. and W.~Chen (2004).
\newblock Sequential optimization and reliability assessment method for
  efficient probabilistic design.
\newblock {\em J. Mech. Design\/}~{\em 126\/}(2), 225--233.

\bibitem[\protect\citeauthoryear{Dubourg}{Dubourg}{2011}]{DubourgThesis}
Dubourg, V. (2011).
\newblock {\em Adaptive surrogate models for reliability analysis and
  reliability-based design optimization}.
\newblock Ph.\ D. thesis, Universit\'e Blaise Pascal, Clermont-Ferrand, France.

\bibitem[\protect\citeauthoryear{Dubourg, Sudret, and Bourinet}{Dubourg
  et~al.}{2011}]{Dubourg2011}
Dubourg, V., B.~Sudret, and J.-M. Bourinet (2011).
\newblock Reliability-based design optimization using {Kriging} and subset
  simulation.
\newblock {\em Struct. Multidisc. Optim.\/}~{\em 44\/}(5), 673--690.

\bibitem[\protect\citeauthoryear{Enevoldsen and S{\o}rensen}{Enevoldsen and
  S{\o}rensen}{1994}]{Enevoldsen1994}
Enevoldsen, I. and J.~D. S{\o}rensen (1994).
\newblock Reliability-based optimization in structural engineering.
\newblock {\em Structural Safety\/}~{\em 15\/}(3), 169--196.

\bibitem[\protect\citeauthoryear{Foschi, Li, and Zhang}{Foschi
  et~al.}{2002}]{Foschi2002}
Foschi, R.~O., H.~Li, and J.~Zhang (2002).
\newblock Reliability and performance-based design: a computational approach
  and applications.
\newblock {\em Structural Safety\/}~{\em 24\/}(2--4), 205--218.

\bibitem[\protect\citeauthoryear{Frangopol}{Frangopol}{1985}]{Frangopol1985}
Frangopol, D.~M. (1985).
\newblock Structural optimization using reliability concepts.
\newblock {\em J. Struct. Eng.\/}~{\em 111\/}(11), 2288--2301.

\bibitem[\protect\citeauthoryear{Frangopol and Maute}{Frangopol and
  Maute}{2003}]{Frangopol2003}
Frangopol, D.~M. and K.~Maute (2003).
\newblock Life-cycle reliability-based optimization of civil and aerospace
  structures.
\newblock {\em Comput. Struct.\/}~{\em 81}, 397--410.

\bibitem[\protect\citeauthoryear{Gao and Li}{Gao and Li}{2017}]{Gao2017}
Gao, T. and J.~Li (2017).
\newblock A derivative-free trust-region algorithm for reliability-based
  optimization.
\newblock {\em Struct. Multidisc. Optim.\/}~{\em 55\/}(4), 1535--1539.

\bibitem[\protect\citeauthoryear{Gaspar, Teixeira, and Guedes~Soares}{Gaspar
  et~al.}{2017}]{Gaspar2017}
Gaspar, B., A.~P. Teixeira, and C.~Guedes~Soares (2017).
\newblock Adaptive surrogate model with active refinement combining {K}riging
  and a trust region method.
\newblock {\em Reliab. Eng. Syst. Saf.\/}~{\em 165}, 277--291.

\bibitem[\protect\citeauthoryear{Geyer, Papaiannou, and Straub}{Geyer
  et~al.}{2019}]{Geyer2019}
Geyer, S., I.~Papaiannou, and D.~Straub (2019).
\newblock Cross-entropy-based importance sampling using gaussian densities
  revisited.
\newblock {\em Structural Safety\/}~{\em 76}, 15--27.

\bibitem[\protect\citeauthoryear{Hansen and Ostermeier}{Hansen and
  Ostermeier}{2001}]{Hansen2001}
Hansen, N. and A.~Ostermeier (2001).
\newblock Completely derandomized self-adaptation in evolution strategies.
\newblock {\em Evol. Comput.\/}~{\em 9\/}(2), 159--195.

\bibitem[\protect\citeauthoryear{Hilton and Feigen}{Hilton and
  Feigen}{1960}]{Hilton1960}
Hilton, H.~H. and M.~Feigen (1960).
\newblock Minimum weight analysis based on structural reliability.
\newblock {\em J. aerospace sci.\/}~{\em 27\/}(9), 641--652.

\bibitem[\protect\citeauthoryear{Jia and Taflanidis}{Jia and
  Taflanidis}{2013}]{Jia2013}
Jia, G. and A.~A. Taflanidis (2013).
\newblock Non-parametric stochastic subset optimization for optimal-reliability
  design problems.
\newblock {\em Comput. Struct.\/}~{\em 126}, 86--99.

\bibitem[\protect\citeauthoryear{Jiang, Qiu, Gao, Cai, and Li}{Jiang
  et~al.}{2017}]{Jiang2017}
Jiang, C., H.~Qiu, L.~Gao, X.~Cai, and P.~Li (2017).
\newblock An adaptive hybrid single-loop method for reliability-based design
  optimization using iterative control strategy.
\newblock {\em Struct. Multidisc. Optim.\/}, 1--16.

\bibitem[\protect\citeauthoryear{Kaveh and Talatahari}{Kaveh and
  Talatahari}{2009}]{Kaveh2009}
Kaveh, A. and S.~Talatahari (2009).
\newblock Particle swarm optimizer, and colony strategy and harmony search
  scheme hybridized for optimization of truss structures.
\newblock {\em Comput. Struct.\/}~{\em 87}, 1267--283.

\bibitem[\protect\citeauthoryear{Kaymaz}{Kaymaz}{2007}]{Kaymaz2007}
Kaymaz, I. (2007).
\newblock Approximation methods for reliability-based design optimization
  problems.
\newblock {\em {GAMM-Mitt}\/}~{\em 30\/}(2), 225--268.

\bibitem[\protect\citeauthoryear{Kharmanda, Mohamed, and Lemaire}{Kharmanda
  et~al.}{2002}]{Kharmanda2002}
Kharmanda, G., A.~Mohamed, and M.~Lemaire (2002).
\newblock Efficient reliability-based design optimization using a hybrid space
  with application to finite element analysis.
\newblock {\em Struct. Multidisc. Optim.\/}~{\em 24\/}(3).

\bibitem[\protect\citeauthoryear{Kurtz and Song}{Kurtz and
  Song}{2013}]{Kurtz2013}
Kurtz, N. and J.~Song (2013).
\newblock Cross-entropy-based adaptive importance sampling using gaussian
  mixture.
\newblock {\em Structural Safety\/}~{\em 42}, 35--44.

\bibitem[\protect\citeauthoryear{Kuschel and Rackwitz}{Kuschel and
  Rackwitz}{1997}]{Kuschel1997}
Kuschel, N. and R.~Rackwitz (1997).
\newblock Two basic problems in reliability-based structural optimization.
\newblock {\em Math. Method. Oper. Res.\/}~{\em 46\/}(3), 309--333.

\bibitem[\protect\citeauthoryear{Lataniotis, Marelli, and Sudret}{Lataniotis
  et~al.}{2018}]{Lataniotis2018}
Lataniotis, C., S.~Marelli, and B.~Sudret (2018).
\newblock The gaussian process modeling module in {UQLab}.
\newblock {\em Soft Computing in Civil Engineering\/}~{\em 2\/}(3), 91--116.

\bibitem[\protect\citeauthoryear{Lee, Choi, Du, and Gorsich}{Lee
  et~al.}{2008}]{Lee2008b}
Lee, I., K.~K. Choi, L.~Du, and D.~Gorsich (2008).
\newblock {Inverse analysis method using MPP-based dimension reduction for
  reliability-based design optimization of nonlinear and multi-dimensional
  systems}.
\newblock {\em Comput. Methods Appl. Mech. Engrg.\/}~{\em 198}, 14--27.

\bibitem[\protect\citeauthoryear{Lee, Choi, and Zhao}{Lee
  et~al.}{2011}]{Lee2011}
Lee, I., K.~K. Choi, and L.~Zhao (2011).
\newblock Sampling-based {RBDO} using the stochastic sensitivity analysis and
  dynamic {K}riging method.
\newblock {\em Struct. Multidisc. Optim.\/}~{\em 44\/}(3), 299--317.

\bibitem[\protect\citeauthoryear{Lee, Yang, and Ruy}{Lee
  et~al.}{2002}]{Lee2002}
Lee, J.-O., Y.-S. Yang, and W.-S. Ruy (2002).
\newblock A comparative study of reliability-index and target-performance-based
  probabilistic structural design optimization.
\newblock {\em Comput. Struct.\/}~{\em 80}, 257--269.

\bibitem[\protect\citeauthoryear{Lee}{Lee}{1997}]{Lee}
Lee, P.~M. (1997).
\newblock {\em Bayesian statistics An introduction, second edition}.
\newblock Arnold.

\bibitem[\protect\citeauthoryear{Lee and Jung}{Lee and Jung}{2008}]{Lee2008}
Lee, T. and J.~Jung (2008).
\newblock {A sampling technique enhancing accuracy and efficiency of
  metamodel-based RBDO: Constraint boundary sampling}.
\newblock {\em Comput. Struct.\/}~{\em 86\/}(13-14), 1463--1476.

\bibitem[\protect\citeauthoryear{Lehk\'y, Slowik, and Nov\'ak}{Lehk\'y
  et~al.}{2017}]{Lekhy2017}
Lehk\'y, D., O.~Slowik, and D.~Nov\'ak (2017).
\newblock Reliability-based design: {A}rtificial neural networks and
  double-loop reliability-based optimization approaches.
\newblock {\em Adv. Eng. Soft.\/}, 1--13.

\bibitem[\protect\citeauthoryear{Li, Meng, and Hu}{Li et~al.}{2015}]{Li2015}
Li, G., Z.~Meng, and H.~Hu (2015).
\newblock An adaptive hybrid approach for reliability-based design
  optimization.
\newblock {\em Struct. Multidisc. Optim.\/}~{\em 51\/}(5), 1051--1065.

\bibitem[\protect\citeauthoryear{Li and Yang}{Li and Yang}{1994}]{Li1994}
Li, W. and L.~Yang (1994).
\newblock An effective optimization procedure based on structural reliability.
\newblock {\em Comput. Struct.\/}~{\em 52\/}(5), 1061--1067.

\bibitem[\protect\citeauthoryear{Li, Qiu, Chen, Gao, and Shao}{Li
  et~al.}{2016}]{Li2016}
Li, X., H.~Qiu, Z.~Chen, L.~Gao, and X.~Shao (2016).
\newblock {A local Kriging approximation method using MPP for reliability-based
  design optimization}.
\newblock {\em Comput. Struct.\/}~{\em 162}, 102--115.

\bibitem[\protect\citeauthoryear{Liang, Mourelatos, and Tu}{Liang
  et~al.}{2004}]{Liang2004}
Liang, J., Z.~Mourelatos, and J.~Tu (2004).
\newblock A single-loop method for reliability-based design optimization.
\newblock In {\em {Proc. DETC'04 ASME 2004 Design engineering technical
  conferences and computers and information in engineering conference, Sept.28
  - Oct. 2, 2004, Salt Lake City, Utah, USA}}.

\bibitem[\protect\citeauthoryear{Liang, Mourelatos, and Nikolaidis}{Liang
  et~al.}{2007}]{Liang2007}
Liang, J., Z.~P. Mourelatos, and E.~Nikolaidis (2007).
\newblock A single-loop approach for system reliability-based design
  optimization.
\newblock {\em J. Mech. Des.\/}~{\em 129\/}(12), 1215 -- 1224.

\bibitem[\protect\citeauthoryear{Lim and Lee}{Lim and Lee}{2016}]{Lim2016}
Lim, J. and B.~Lee (2016).
\newblock A semi-single-loop method using approximatino of most probable point
  for reliability-based design optimization.
\newblock {\em Struct. Multidisc. Optim.\/}~{\em 53\/}(4), 745--757.

\bibitem[\protect\citeauthoryear{Liu and Cheung}{Liu and
  Cheung}{2017}]{Liu2017}
Liu, W.-S. and S.~H. Cheung (2017).
\newblock Reliability based design optimization with approximate failure
  probability function in partitioned design space.
\newblock {\em Reliab. Eng. Syst. Saf.\/}~{\em 167}, 602--611.

\bibitem[\protect\citeauthoryear{Madsen and Hansen}{Madsen and
  Hansen}{1992}]{Madsen1992}
Madsen, H.~O. and P.~F. Hansen (1992).
\newblock A comparison of some algorithms for reliability based structural
  optimization and sensitivity analysis.
\newblock In R.~Rackwitz and P.~Thoft-Christensen (Eds.), {\em Reliability and
  Optimization of Structural Systems'91. Lectures Notes in Engineering},
  Volume~76, pp.\  443--451. Springer, Berlin, Heidelberg.

\bibitem[\protect\citeauthoryear{Marelli and Sudret}{Marelli and
  Sudret}{2014}]{MarelliUQLab2014}
Marelli, S. and B.~Sudret (2014).
\newblock {UQLab}: A framework for uncertainty quantification in {Matlab}.
\newblock In {\em Vulnerability, Uncertainty, and Risk (Proc. 2nd Int. Conf. on
  Vulnerability, Risk Analysis and Management {(ICVRAM2014)}, Liverpool, United
  Kingdom)}, pp.\  2554--2563.

\bibitem[\protect\citeauthoryear{McKay, Beckman, and Conover}{McKay
  et~al.}{1979}]{McKay1979}
McKay, M.~D., R.~J. Beckman, and W.~J. Conover (1979).
\newblock A comparison of three methods for selecting values of input variables
  in the analysis of output from a computer code.
\newblock {\em Technometrics\/}~{\em 2}, 239--245.

\bibitem[\protect\citeauthoryear{Moustapha, Lataniotis, Marelli, and
  Sudret}{Moustapha et~al.}{2018}]{UQdoc_11_111}
Moustapha, M., C.~Lataniotis, S.~Marelli, and B.~Sudret (2018).
\newblock {UQLab} user manual -- {S}upport vector machines for regression.
\newblock Technical report, Chair of Risk, Safety \& Uncertainty
  Quantification, ETH Zurich.
\newblock Report \# UQLab-V1.1-111.

\bibitem[\protect\citeauthoryear{Moustapha, Marelli, and Sudret}{Moustapha
  et~al.}{2019}]{UQdoc_12_114}
Moustapha, M., S.~Marelli, and B.~Sudret (2019).
\newblock {UQLab} user manual -- {R}eliability-based design optimization.
\newblock Technical report, Chair of Risk, Safety \& Uncertainty
  Quantification, ETH Zurich.
\newblock Report \# UQLab-V1.2-114.

\bibitem[\protect\citeauthoryear{Moustapha and Sudret}{Moustapha and
  Sudret}{2017}]{MoustaphaICOSSAR2017}
Moustapha, M. and B.~Sudret (2017).
\newblock Quantile-based optimization under uncertainties using bootstrap
  polynomial chaos expansions.
\newblock In {\em Proc. 12th~Internatinoal Conference on Structural Safety and
  Reliability (ICOSSAR), August 6-10, 2017, Vienna, Austria}.

\bibitem[\protect\citeauthoryear{Moustapha, Sudret, Bourinet, and
  Guillaume}{Moustapha et~al.}{2016}]{MoustaphaSMO2016}
Moustapha, M., B.~Sudret, J.-M. Bourinet, and B.~Guillaume (2016).
\newblock Quantile-based optimization under uncertainties using adaptive
  {K}riging surrogate models.
\newblock {\em Struct. Multidisc. Optim.\/}~{\em 54\/}(6), 1403--1421.

\bibitem[\protect\citeauthoryear{Moustapha, Sudret, Bourinet, and
  Guillaume}{Moustapha et~al.}{2018}]{MoustaphaJRUES2018}
Moustapha, M., B.~Sudret, J.-M. Bourinet, and B.~Guillaume (2018).
\newblock Comparative study of {K}riging and support vector regression for
  structural engineering applications.
\newblock {\em ASCE-ASME J. Risk Uncertainty Eng. Syst., Part A: Civ.
  Eng.\/}~{\em 4\/}(2).

\bibitem[\protect\citeauthoryear{Nikolaidis and Burdisso}{Nikolaidis and
  Burdisso}{1988}]{Nikolaidis1988}
Nikolaidis, E. and R.~Burdisso (1988).
\newblock Reliability based optimization: {A} safety index approach.
\newblock {\em Comput. Struct.\/}~{\em 28\/}(6), 781--788.

\bibitem[\protect\citeauthoryear{Papadrakakis, Lagaros, and
  Plevris}{Papadrakakis et~al.}{2005}]{Papadrakakis2005}
Papadrakakis, M., N.~D. Lagaros, and V.~Plevris (2005).
\newblock Design optimization of steel structures considering uncertainties.
\newblock {\em Eng. Struct.\/}~{\em 27\/}(9), 1408--1418.

\bibitem[\protect\citeauthoryear{Papaioannou, Betz, Zwirglmaier, and
  Straub}{Papaioannou et~al.}{2015}]{Papaioannou2015}
Papaioannou, I., W.~Betz, K.~Zwirglmaier, and D.~Straub (2015).
\newblock {MCMC} algorithms for subset simulation.
\newblock {\em Prob. Eng. Mech.\/}~{\em 41}, 89 -- 103.

\bibitem[\protect\citeauthoryear{Pradlwarter, Schu\"eller, Koutsourelakis, and
  Charmpis}{Pradlwarter et~al.}{2007}]{Pradlwarter2007}
Pradlwarter, H.~J., G.~I. Schu\"eller, P.~S. Koutsourelakis, and D.~C. Charmpis
  (2007).
\newblock Application of line sampling simulation method to reliability
  benchmark problems.
\newblock {\em Structural Safety\/}~{\em 29\/}(3), 208--221.

\bibitem[\protect\citeauthoryear{Rahman and Wei}{Rahman and
  Wei}{2008}]{Rahman2008b}
Rahman, S. and D.~Wei (2008).
\newblock Design sensitivity and reliability-based structural optimization by
  univariate decomposition.
\newblock {\em Struct. Multidisc. Optim.\/}~{\em 35\/}(3), 245--261.

\bibitem[\protect\citeauthoryear{Rahman and Xu}{Rahman and
  Xu}{2004}]{Rahman2004}
Rahman, S. and H.~Xu (2004).
\newblock A univariate dimension-reduction method for multi-dimensional
  integration in stochastic mechanics.
\newblock {\em Prob. Eng. Mech.\/}~{\em 19}, 393--408.

\bibitem[\protect\citeauthoryear{Rashki, Miri, and Moghaddam}{Rashki
  et~al.}{2014}]{Rashki2014}
Rashki, M., M.~Miri, and M.~A. Moghaddam (2014).
\newblock A simulation-based method for reliability-based design optimization
  problems with highly nonlinear constraints.
\newblock {\em Automation in Construction\/}~{\em 47}, 24--36.

\bibitem[\protect\citeauthoryear{Rasmussen and Williams}{Rasmussen and
  Williams}{2006}]{Rasmussen2006}
Rasmussen, C.~E. and C.~K.~I. Williams (2006).
\newblock {\em Gaussian processes for machine learning\/} ({Internet} ed.).
\newblock Adaptive computation and machine learning. Cambridge, Massachusetts:
  MIT Press.

\bibitem[\protect\citeauthoryear{Royset}{Royset}{2004}]{Royset2004}
Royset, J.~O. (2004).
\newblock Reliability-based optimal design using sample average approximations.
\newblock {\em Prob. Eng. Mech.\/}~{\em 19}, 331--343.

\bibitem[\protect\citeauthoryear{Royset, {Der Kiureghian, A.}, and
  Polak}{Royset et~al.}{2001}]{Royset2001}
Royset, J.~O., {Der Kiureghian, A.}, and E.~Polak (2001).
\newblock Reliability-based optimal structural design by the decoupling
  approach.
\newblock {\em Reliab. Eng. Sys. Safety\/}~{\em 73\/}(3), 213-- 221.

\bibitem[\protect\citeauthoryear{Santner, Williams, and Notz}{Santner
  et~al.}{2003}]{Santner2003}
Santner, T.~J., B.~J. Williams, and W.~I. Notz (2003).
\newblock {\em The {D}esign and {A}nalysis of {C}omputer {E}xperiments}.
\newblock Springer, New York.

\bibitem[\protect\citeauthoryear{Shetty, Guedes-Soares, Thoft-Christensen, and
  Jensen}{Shetty et~al.}{1998}]{Shetty1998}
Shetty, N.~K., C.~Guedes-Soares, P.~Thoft-Christensen, and F.~M. Jensen (1998).
\newblock Fire safety assessment and optimal design of passive fire protection
  for offshore structures.
\newblock {\em Reliab. Eng. Syst. Saf.\/}~{\em 61\/}(1-2), 139--149.

\bibitem[\protect\citeauthoryear{Smola and Sch\"olkopf}{Smola and
  Sch\"olkopf}{2004}]{Smola2004}
Smola, A.~J. and B.~Sch\"olkopf (2004).
\newblock A tutorial on support vector regression.
\newblock {\em Stat. Comput.\/}~{\em 14}, 199--222.

\bibitem[\protect\citeauthoryear{{Sobol'}}{{Sobol'}}{1967}]{Sobol1967}
{Sobol'}, I.~M. (1967).
\newblock Distribution of points in a cube and approximate evaluation of
  integrals.
\newblock {\em U.S.S.R Comput. Maths. Math. Phys.\/}~{\em 7}, 86--112.

\bibitem[\protect\citeauthoryear{Song}{Song}{2013}]{SongThesis2013}
Song, H. (2013).
\newblock {\em Efficient sampling-based {RBDO} by using virtual support vector
  machine and improving the accuracy of the {K}riging method}.
\newblock Ph.\ D. thesis, University of Iowa, USA.

\bibitem[\protect\citeauthoryear{Spall}{Spall}{1998a}]{Spall1998b}
Spall, J.~C. (1998a).
\newblock Implementation of the simultaneous perturbation algorithm for
  stochastic optimization.
\newblock {\em IEEE Trans. Aerospace Electronic Systems\/}~{\em 34\/}(3),
  817--823.

\bibitem[\protect\citeauthoryear{Spall}{Spall}{1998b}]{Spall1998a}
Spall, J.~C. (1998b).
\newblock An overview of the simultaneous perturbation method for efficient
  optimization.
\newblock {\em Johns Hopkins Apl. Technical Digest\/}~{\em 19\/}(4), 482--492.

\bibitem[\protect\citeauthoryear{Spall}{Spall}{2003a}]{Spall2003_ch14}
Spall, J.~C. (2003a).
\newblock {\em Introduction to stochastic search and optimization:
  {E}stimation, simulation and control}, Chapter 14: {S}imulation-based
  optimization {I}: regression, common random numbers, and selection methods.
\newblock John Wiley \& Sons.

\bibitem[\protect\citeauthoryear{Spall}{Spall}{2003b}]{Spall2003}
Spall, J.~C. (2003b).
\newblock {\em Introduction to stochastic search and optimization:
  {E}stimation, simulation and control}.
\newblock John Wiley \& Sons.

\bibitem[\protect\citeauthoryear{Str\"omberg}{Str\"omberg}{2017}]{Stroemberg2017}
Str\"omberg, N. (2017).
\newblock Reliability-based design optimization using {SORM} and {SQP}.
\newblock {\em Struct. Multidisc. Optim.\/}~{\em 56\/}(3), 631--645.

\bibitem[\protect\citeauthoryear{Taflanidis}{Taflanidis}{2007}]{TaflanidisThesis2007}
Taflanidis, A.~A. (2007).
\newblock {\em Stochastic system design and applications to stochastic robust
  structural control}.
\newblock Ph.\ D. thesis, California Institute of Technology, Pasadena,
  California, USA.

\bibitem[\protect\citeauthoryear{Taflanidis and Beck}{Taflanidis and
  Beck}{2008}]{Taflanidis2008}
Taflanidis, A.~A. and J.~L. Beck (2008).
\newblock Stochastic subset optimization for optimal reliability problems.
\newblock {\em Prob. Eng. Mech\/}~{\em 23}, 324--338.

\bibitem[\protect\citeauthoryear{Taflanidis and Medina}{Taflanidis and
  Medina}{2014}]{Taflanidis2014}
Taflanidis, A.~J. and A.~C. Medina (2014).
\newblock Adaptive {K}riging for simulation-based design under uncertainty:
  {D}evelopment of metamodels in augmented input space and adaptive tuning of
  their characteristics.
\newblock In {\em {Proc. 4th International Conference On Simulation And
  Modeling Methodologies, Technologies And Applications, August 28-30, 2014,
  Vienna, Austria}}.

\bibitem[\protect\citeauthoryear{Torre, Marelli, Embrechts, and Sudret}{Torre
  et~al.}{2018}]{Torre2018_Vines4UQ}
Torre, E., S.~Marelli, P.~Embrechts, and B.~Sudret (2018).
\newblock A general framework for data-driven uncertainty quantification under
  complex input dependencies using vine copulas.
\newblock {\em Prob. Eng. Mech.\/}.
\newblock (in press).

\bibitem[\protect\citeauthoryear{Tu, Choi, and Park}{Tu et~al.}{1999}]{Tu1999}
Tu, J., K.~K. Choi, and Y.~H. Park (1999).
\newblock A new study on reliability-based design optimization.
\newblock {\em J. Mech. Des.\/}~{\em 121}, 557 -- 564.

\bibitem[\protect\citeauthoryear{Valdebenito and Schu\"eller}{Valdebenito and
  Schu\"eller}{2010}]{Valdebenito2010}
Valdebenito, A.~M. and G.~I. Schu\"eller (2010).
\newblock A survey on approaches for reliability-based optimization.
\newblock {\em Struct. Multidisc. Optim.\/}~{\em 42}, 645--663.

\bibitem[\protect\citeauthoryear{Vapnik}{Vapnik}{1995}]{Vapnik:1995}
Vapnik, V.~N. (1995).
\newblock {\em The {N}ature of {S}tatistical {L}earning {T}heory}.
\newblock Springer-Verlag, New York.

\bibitem[\protect\citeauthoryear{Wang}{Wang}{2003}]{Wang2003}
Wang, G.~G. (2003).
\newblock Adaptive response surface method using inherited {L}atin {H}ypercube
  design points.
\newblock {\em J. Mech. Design\/}~{\em 125}, 210--220.

\bibitem[\protect\citeauthoryear{Wang}{Wang}{1998}]{Wang1998}
Wang, I.-J. ad~Spall, J.~C. (1998).
\newblock A constrained simulation perturbation stochastic approximation
  algorithm based on penalty functions.
\newblock In {\em Proceedings of the 1998 IEEE ISIC/CIRA/ISAS Joint Conference,
  Sept. 14--17, 1998, Gaithersburg, MD, USA}.

\bibitem[\protect\citeauthoryear{Youn}{Youn}{2007}]{Youn2007}
Youn, B.~D. (2007).
\newblock Adaptive-loop method for non-deterministic design optimization.
\newblock {\em {Proceedings of the Institution of Mechanical Engineers, Part O:
  Journal of Risk and Reliability}\/}~{\em 221\/}(2), 107--116.

\bibitem[\protect\citeauthoryear{Youn, Choi, and Du}{Youn
  et~al.}{2005}]{Youn2005}
Youn, B.~D., K.~K. Choi, and L.~Du (2005).
\newblock Enriched performance measure approach for reliability-based design
  optimization.
\newblock {\em AIAA Journal\/}~{\em 43\/}(4), 874--884.

\bibitem[\protect\citeauthoryear{Zhang, Taflanidis, and Medina}{Zhang
  et~al.}{2017}]{Zhang2017}
Zhang, J., A.~A. Taflanidis, and J.~C. Medina (2017).
\newblock {Sequential approximate optimization for design under uncertainty
  problems utilizing Kriging metamodeling in augmented input space}.
\newblock {\em Comput. Methods Appl. Mech. Engrg.\/}~{\em 315}, 369--395.

\bibitem[\protect\citeauthoryear{Zou and Mahadevan}{Zou and
  Mahadevan}{2006}]{Zou2006}
Zou, T. and S.~Mahadevan (2006).
\newblock A direct decoupling approach for efficient reliability-based design
  optimization.
\newblock {\em Struct. Multidisc. Optim.\/}~{\em 31}, 190--200.

\end{thebibliography}
\end{document}